\documentclass[preprintnumbers,aps,11pt,superscriptaddress,nofootinbib,tightenlines,floatfix]{revtex4}
\usepackage{amsmath,amssymb}
\usepackage{graphicx}
\usepackage{bm}
\usepackage{comment}
\usepackage{color}
\usepackage{subfigure}
\usepackage{array}
\usepackage{multirow}
\usepackage{natbib}
\usepackage{url}
\usepackage[]{units}

\usepackage[T1]{fontenc}
\usepackage[latin9]{inputenc}
\usepackage{graphicx}
\usepackage{esint}
\usepackage{hyperref}
\usepackage{atbegshi}
\usepackage{lipsum}

\def\si{^1 \hskip -0.025in S _0}
\def\siii{^3 \hskip -0.025in S _1}
\def\diii{^3 \hskip -0.025in D _1}

\begin{document}

\dimen\footins=5\baselineskip\relax

\preprint{\vbox{
\hbox{INT-PUB-13-045}
\hbox{JLAB-THY-13-1827} 
}}

\title{Two-Baryon Systems with Twisted Boundary Conditions
}
\author{Ra\'ul A. Brice\~no{\footnote{\tt rbriceno@jlab.org}}}
\affiliation{Jefferson Laboratory, 12000 Jefferson Avenue, Newport
  News, VA 23606, USA}

\author{Zohreh Davoudi{\footnote{\tt davoudi@uw.edu}}
}
\affiliation{Department of Physics, University of Washington,
 Box 351560, Seattle, WA 98195, USA}
\affiliation{Institute for Nuclear Theory, Box 351550, Seattle, WA 98195-1550, USA}

\
\author{Thomas C. Luu{\footnote{\tt t.luu@fz-juelich.de}}
}
\affiliation{Institute for Advanced Simulation,
Forschungszentrum J\"{u}lich, D--52425 J\"{u}lich, Germany}
\affiliation{Institut f\"ur Kernphysik and J\"ulich Center for Hadron Physics,
Forschungszentrum J\"{u}lich, D--52425 J\"{u}lich, Germany}

\author{Martin J. Savage{\footnote{\tt mjs5@uw.edu}}}
\affiliation{Department of Physics, University of Washington,
 Box 351560, Seattle, WA 98195, USA}
\affiliation{Institute for Nuclear Theory, Box 351550, Seattle, WA 98195-1550, USA}

\date{\today}

\begin{abstract} 
We explore the use of twisted boundary conditions in extracting the nucleon mass and the binding energy of two-baryon systems, 
such as the deuteron, from Lattice QCD calculations.  
Averaging the results of calculations performed with periodic and anti-periodic boundary conditions imposed 
upon the light-quark fields,
or other pair-wise averages,
improves the volume dependence of the deuteron binding energy from 
$\sim e^{- \kappa{\rm L}}/{\rm L} $ to $\sim e^{- \sqrt{2} \kappa{\rm L}}/{\rm L}$.
However,  a twist angle of $\pi/2$ in each of the spatial directions
improves the volume dependence
from $\sim e^{- \kappa{\rm L}}/{\rm L} $ to $\sim e^{- 2 \kappa{\rm L}}/{\rm L}$.
Twist averaging the binding energy with a random sampling of twist angles 
improves the volume dependence
from $\sim e^{- \kappa{\rm L}}/{\rm L} $ to $\sim e^{- 2 \kappa{\rm L}}/{\rm L}$,
but with a standard deviation of $\sim e^{- \kappa{\rm L}}/{\rm L} $,
introducing a signal-to-noise issue in modest lattice volumes.
Using the experimentally determined phase shifts and mixing angles, we determine the expected 
energies of the deuteron states over a range of cubic lattice volumes for a selection of twisted boundary conditions.

\end{abstract}
\maketitle

%%%%%%%%%%%%%%%%%%%%%%%%%%%%%%%%%%%%%%%%%%%
\section{Introduction  
\label{sec:Intro} 
}
\noindent
Lattice quantum chromodynamics (LQCD) 
is evolving into a quantitative tool with which to describe the low-energy dynamics of 
few-body hadronic systems. 
After fully quantifying both the statistical and systematic uncertainties that are inherent in LQCD calculations,
the masses of the lowest-lying hadrons are found to be in impressive agreement with those of nature~\cite{Kronfeld:2012uk, 2012arXiv1209.3468K}.
Recently, the ground-state energies of the s-shell nuclei and 
hypernuclei
have been determined at a small number of light-quark masses~\cite{Beane:2010hg,Beane:2011zpa,Inoue:2010es,Inoue:2011ai,
Beane:2011iw, Beane:2012vq, Yamazaki:2012hi, Yamazaki:2013rna}.
Through algorithmic improvements, along with the growth in available computational resources,
such calculations are moving towards the physical values of quark masses, 
and to calculations of nuclear properties such as magnetic moments.
It is exciting to realize that 
within the next few years, LQCD calculations will provide a firm foundation for the 
forces between nucleons directly from QCD, e.g. Ref.~\cite{Barnea:2013uqa}.
One of the few systematic uncertainties present in the results of LQCD calculations 
arises in the infinite-volume extrapolation from finite-spacetime lattices,
which, in fact,  can be quantified by performing calculations in a range of lattice volumes. 
For simple systems such as $\pi^+\pi^+$~\cite{Beane:2011sc,Dudek:2012gj}, 
using effective field theory (EFT) methods, 
or direct knowledge of the S-matrix, 
the functional volume dependence of observables is available
to provide well-defined predictions in infinite volume.

The finite-volume (FV) corrections to the mass of hadrons are dominated by the pion mass, $m_{\pi}$,
and for a cubic volume with the spatial extent ${\rm L}$ 
the leading order (LO) corrections scale as 
$e^{-m_{\pi}{\rm L}}/{\rm L}$~\cite{Luscher:1985dn}. 
For two-body bound states,
the size of the bound state provides a second scale
responsible for volume modifications.
These scale as 
$e^{-\kappa {\rm L}}/{\rm L}$  at LO in the volume expansion~\cite{Luscher:1990ux, Beane:2003da}, 
where
$\kappa$ is the binding momentum of the bound state comprised of two particles of masses 
$m_1$ and $m_2$ with a binding energy of $B=-(\sqrt{-\kappa^2+m_1^2}+\sqrt{-\kappa^2+m_2^2}-m_1-m_2)$. 
For the deuteron, 
which is the only two-nucleon bound state in nature and is  bound by only 
$B_d^{\infty}=2.224644(34)~{\rm MeV}$, 
these latter volume corrections 
can be large even in a modest lattice volume for generic boundary conditions (BCs) imposed upon the 
quark fields.
As an example, 
an extraction of the deuteron binding energy that is accurate at the percent level 
from quark fields subject to periodic boundary conditions (PBCs) 
requires volumes with ${\rm L}\gtrsim17~{\rm fm}$.
However,  this is the worst-case scenario  as, for instance, 
also producing correlation functions corresponding to a non-zero center-of-mass (CM) momentum
allows for an exponential reduction in the volume dependence~\cite{Davoudi:2011md, Briceno:2013bda}.
Such calculations will permit single-volume determinations of the deuteron binding energy 
with percent-level accuracy 
in significantly smaller volumes. 

LQCD calculations are commonly performed with PBCs imposed upon the quark fields in the spatial directions,
constraining the quark momentum modes in the volume to satisfy
$\mathbf{p}=\frac{2\pi}{L}\mathbf{n}$ with $\mathbf{n}$ being an integer triplet. 
PBCs are a subset of a larger class of BCs called twisted BCs (TBCs).
 TBCs~\cite{PhysRevLett.7.46} are those that require the quark fields to acquire 
a  phase $\theta$ at the boundary, 
$\psi(\mathbf{x}+\mathbf{n}{\rm{L}})=e^{i{\rm{\theta}} \cdot \mathbf{n}}\psi(\mathbf{x})$,
where $0<\theta_i<2\pi$ is the twist angle in the $i^{\rm th}$ Cartesian direction. 
Bedaque~\cite{Bedaque:2004kc}  introduced this idea to the LQCD community, and showed  that 
TBCs are equivalent to having a $U(1)$ background gauge field in the QCD Lagrangian 
with the quarks subject to PBCs. 
An arbitrary momentum can be selected for a (non-interacting) hadron by a judicious choice
of the twist angles of its valence quarks, 
$\mathbf{p}=\frac{2\pi}{L}\mathbf{n}+\frac{\bm{\phi}}{L}$,
where $\bm{\phi}$ is the sum of the twists of the valence quarks, again with 
$0<\phi_i<2\pi$, and $\mathbf{n}$ is an integer triplet. 
TBCs have been  shown to 
be useful in LQCD calculations of 
 the low-momentum transfer behavior of form factors required in determining 
hadron radii and moments, circumventing the need for large-volume lattices~\cite{Tiburzi:2005hg,Jiang:2006gna,Boyle:2007wg,Simula:2007fa,Boyle:2008yd,Aoki:2008gv,Boyle:2012nb, Brandt:2013mb}.
They have also been speculated to be helpful 
in calculations of $K\rightarrow \pi\pi$ decays
by bringing the initial and final FV states  closer in energy~\cite{deDivitiis:2004rf, Sachrajda:2004mi}.

In addition to performing calculations with a particular twist,
by averaging the results of calculations over twist angles, the discrete sum over momentum modes becomes 
an integral over momenta, 
\begin{eqnarray}
\int \ {d^3\bm{\phi}\over (2\pi)^3}\ 
\frac{1}{{\rm L}^3}\ \sum_{\mathbf{n} \in \mathbb{Z}^3}
& \equiv & 
\int \frac{d^3\mathbf{p}}{(2\pi)^3}
\ \ \ . 
\end{eqnarray}
Although the volume dependence of most quantities   is non linear due to interactions, such averaging can eliminate significant FV effects. 
This was first examined in the context of condensed-matter physics where, for example, the finite-size effects in the finite-cluster calculations 
of correlated electron systems are shown to be reduced by the boundary condition integration technique \cite{Gros, PhysRevB.53.6865}. 
This technique is implemented in quantum Monte Carlo (QMC) algorithms of many-body  systems, and results in faster convergence of energies to the thermodynamic limit~\cite{2001PhRvE..64a6702L}. 

In this paper, we discuss the advantages of using TBCs to reduce the FV modifications to the mass of hadrons and to the 
binding energy of two-hadron bound states, such as the deuteron. 
In particular, we consider the FV effects resulting 
from averaging the results obtained from PBC and anti-PBCs (APBCs), 
from a specific choice of the twist angle,  i-PBCs, 
and from averaging over twist angles.
For the two-nucleon systems, the volume improvement is explored both analytically and numerically with the use of the 
recently developed FV formalism for nucleon-nucleon (NN) systems that is generalized to systems with TBCs. 
As was first noted by Bedaque and Chen~\cite{Bedaque:2004ax},  the need to generate new gauge field configurations 
with fully twisted BCs can be circumvented by  imposing TBCs on the valence quarks only, which defines partial twisting. 
Partial twisting gives rise to corrections beyond full twisting that scale as $e^{-m_{\pi}{\rm L}}/{\rm L}$, 
and can be neglected for sufficiently large volumes compared to the FV effects from the size of weakly  bound states. 
Although the validity of partial twisting makes it feasible to achieve an approximate twist-averaged result in LQCD calculations, 
this remains a computationally expensive technique.
We demonstrate that certain hadronic twist angles can result in an exponentially-improved convergence to the infinite-volume 
limit of certain quantities, with an accuracy that is comparable  to the  twist-averaged mean.
Further, we speculate that  similar improvements are also present in arbitrary n-body systems.

In some situations it is desirable to keep the volume finite as the extraction of physical 
quantities relies on non-vanishing FV effects. 
This is the well-known L\"uscher methodology~\cite{Luscher:1986pf, Luscher:1990ux}, 
where the  
$2\rightarrow2$ elastic scattering amplitude  can be obtained from the discrete energy eigenvalues of the two particles in a FV 
(see also Refs. \cite{Rummukainen:1995vs, Beane:2003da,Detmold:2004qn,Feng:2004ua,Kim:2005gf, Christ:2005gi, 
Liu:2005kr,He:2005ey,Bernard:2008ax, Lage:2009zv, Ishizuka:2009bx,Davoudi:2011md, Fu:2011xz, Bour:2011ef,   
Guo:2012hv, Li:2012bi, Leskovec:2012gb,  Hansen:2012tf, Briceno:2012yi, Polejaeva:2012ut, Briceno:2012rv,
Briceno:2013bda,Hansen:2013dla} 
for various extensions of the L\"uscher formula). 
A prominent example, as discussed in Refs.~\cite{Briceno:2013lba, Briceno:2013bda}, is the 
FV analysis of the two-nucleon system in the $^3{S_1}$-${^3}{D_1}$ coupled channels. 
The ability to extract the $S$-$D$ mixing parameter, $\epsilon_1$,  and consequently the D/S ratio of the deuteron, 
depends upon the  FV modifications to the binding energy when the deuteron is boosted in particular directions 
within the lattice volume \cite{Briceno:2013bda}. 
The use of TBCs will further enhance the effectiveness of such calculations.
By appropriate choices of the twist angles of each hadron, different CM energies can be accessed in a 
single lattice volume,  further constraining the scattering parameters with the use of L\"uscher's method 
(see e.g. Refs.~\cite{Bernard:2010fp, Doring:2011vk, Doring:2012eu, Ozaki:2012ce} for demonstrations of this technique in 
studying hadronic resonances). 
Due to the possibility of partial twisting in NN scattering, 
these extra energy levels can be obtained without having to generate additional ensembles of
gauge-field  configurations, in analogy with the boosted calculations
(this technique has recently been used to calculate $J/\psi$-$\phi$ scattering~\cite{Ozaki:2012ce}).
Of course, the spectra of energy eigenvalues determined with a range of twist angles allow for fits to 
parametrizations of the S-matrix elements, which can then be used to predict infinite-volume quantities, such as binding 
energies~\cite{Beane:2010em,Prelovsek:2013sxa}.
TBCs provide a way to reduce the systematic uncertainties  that are currently present in analyses of 
coupled-channels systems by providing the ability to control, at some level, 
the location of eigenstates.

%%%%%%%%%%%%%%%%%%%%%%%%%%%%%%%%%%%%%%%%%%%%%%%%%%%%%%%%%%
\section{The Nucleon
\label{sec:Single}
}
\noindent
If the up and down quarks have distinct twist angles,  the charged pions, the proton and the neutron will acquire 
net twist angles denoted as 
${\bm{\phi}}^{\pi^{+}}=-\bm{\phi}^{\pi^{-}}$, $\bm{\phi}^{p}$ and $\bm{\phi}^{n}$, respectively, 
while the flavor-singlet mesons, such as $\pi^0$, will remain untwisted, $\bm{\phi}^{\pi^{0}}=\mathbf{0}$.
The optimal set of quark twists depends upon the desired observable, and 
an appropriate choice  can yield a relation between the twists of different hadrons, or
leave a hadron untwisted.

\begin{figure}[!ht]
\begin{center}  
\includegraphics[scale=0.5]{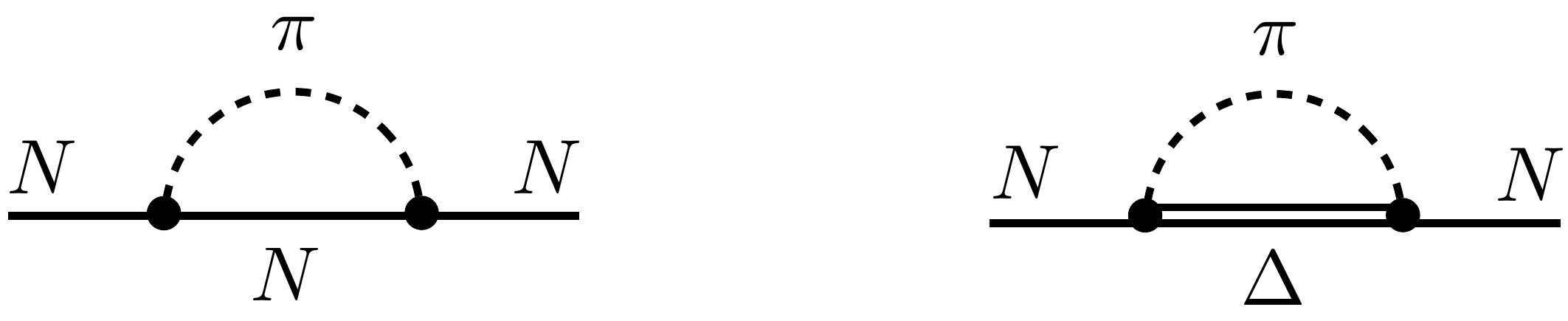}
\caption
{Leading loop contributions to the mass of the nucleon. 
The solid line, solid-double line and dashed line denote a nucleon, a $\Delta$ resonance and a pion, respectively. 
The black disks denote  axial couplings.
}
\label{one-loop}
\end{center}
\end{figure}
The FV corrections to the mass of nucleon $M_N$, in a cubic volume with PBCs imposed on the quark fields, 
have been calculated at one-loop order in two-flavor baryon $\chi$PT \cite{AliKhan:2003cu, Beane:2004tw}
(the three-flavor result can be found in Ref.~\cite{Beane:2011pc}).
Including $\Delta$ resonance as a degree of freedom, these corrections are obtained from the diagrams in 
Fig.~\ref{one-loop}, and take the form~\cite{Beane:2004tw}~\footnote{
Note that we have chosen to define the  $\mathcal{K}(\Delta)$ function with a negative sign compared to
Ref.~\cite{Beane:2004tw}.
}
\begin{eqnarray}
\delta_{{\rm L}} M_{N} \equiv M_N({\rm L})-M_N(\infty)= \frac{3 g_A^2}{8 \pi^2 f_{\pi}^2} \mathcal{K}(0)
+\frac{g_{\Delta N}^2}{3 \pi^2 f_{\pi}^2} \mathcal{K}(\Delta)
\ \ \ \  ,
\label{deltaM}
\end{eqnarray}
where
\begin{eqnarray}
\mathcal{K}(0)= \frac{\pi}{2}m_{\pi}^2 \sum_{\mathbf{n}\neq \mathbf{0}} \frac{e^{-|\mathbf{n}| m_{\pi} {\rm L}}}{|\mathbf{n}| {\rm L}}
\ \ \ \  ,
\end{eqnarray}
and
\begin{eqnarray}
\mathcal{K}(\Delta)
\ =\ 
\int_{0}^{\infty} d \lambda ~ \beta_{\Delta} ~ \sum_{\mathbf{n}\neq \mathbf{0}} 
\left[ 
\beta_{\Delta} K_0(\beta_{\Delta}|\mathbf{n}| {\rm L})
\ -\ 
\frac{1}{|\mathbf{n}| {\rm L}} K_1(\beta_{\Delta}|\mathbf{n}| {\rm L})
\right]
\ \ \ \ \ .
\end{eqnarray}
$m_{\pi}$ and $f_{\pi}$ are the pion mass and decay constant, and $g_{A}$ and $g_{\Delta N}$ denote the nucleon 
axial charge and the $\Delta$-nucleon coupling constant, respectively. $K_n(z)$ is the modified Bessel function of 
the second kind. $\beta_{\Delta} = \lambda^2 + 2 \lambda \Delta + m_{\pi}^2$ where $\Delta$ denotes the 
nucleon-$\Delta$ mass splitting. When expanded in the limit of large L, Eq.~(\ref{deltaM}) scales as $e^{-m_{\pi}{\rm L}}/{\rm L}$ at LO.

Heavy baryon $\chi$PT (HB$\chi$PT) has been used to calculate the masses of the proton and neutron in a FV
at the 
one-loop level with TBCs~\cite{Jiang:2008ja}.~\footnote{The FV corrections to meson masses, decay constants and semileptonic form factors 
with both the TBCs and the partially-TBCs have been calculated at LO in $\chi$PT in Ref. \cite{Sachrajda:2004mi}.
}
The Poisson re-summation formula makes it possible to 
factor the dependences on the twist angles as pure phases, allowing the expressions for the masses to be put into a 
simple form.
The proton mass is found to be 
\begin{eqnarray}
\delta_{{\rm L}} M_{p}=\frac{g_A^2}{4 \pi^2 f_{\pi}^2} \mathcal{K}^{p}(0;\bm{\phi}^{\pi}) + \frac{g_{\Delta N}^2}{6 \pi^2 f_{\pi}^2} 
\mathcal{K}^{p}(\Delta;\bm{\phi}^{\pi})
\ \ \ \ ,
\label{eq:MnucVol}
\end{eqnarray}
where
\begin{eqnarray}
\mathcal{K}^{p}(0;\bm{\phi}^{\pi})= \frac{\pi}{2}m_{\pi}^2 \sum_{\mathbf{n}\neq \mathbf{0}} 
\frac{e^{-|\mathbf{n}| m_{\pi} {\rm L}}}{|\mathbf{n}| {\rm L}} (\frac{1}{2}+e^{-i\mathbf{n}\cdot\bm{\phi}^{\pi^+}})
\ \ \ \ ,
\label{eq:NNpisum}
\end{eqnarray}
and
\begin{eqnarray}
\mathcal{K}^{p}(\Delta;\bm{\phi}^{\pi})
\ =\ 
\int_{0}^{\infty} d \lambda ~ \beta_{\Delta}  
&&\sum_{\mathbf{n}\neq \mathbf{0}} 
\left[ 
\beta_{\Delta} 
K_0(\beta_{\Delta}|\mathbf{n}| {\rm L})
\ -\ 
\frac{1}{|\mathbf{n}| {\rm L}} K_1(\beta_{\Delta}|\mathbf{n}| {\rm L})
\right]
\nonumber\\
&&\qquad \qquad \qquad 
~ \times (e^{-i\mathbf{n}\cdot\bm{\phi}^{\pi^-}}+\frac{2}{3}+\frac{1}{3}e^{-i\mathbf{n}\cdot\bm{\phi}^{\pi^+}})
\ \ \ ,
\label{eq:NDpisum}
\end{eqnarray}
and the neutron mass can be found from these expressions by  the substitutions 
$p\rightarrow n$ and $\pi^+\leftrightarrow\pi^-$.
It is convenient to consider the periodic images associated with the nucleon 
having their contributions modified by the appropriate phase factor 
due to the TBCs.

After twist averaging (over the twists of the pion field, see Appendix~\ref{app: TI}), 
the leading FV corrections to the mass of both the proton 
and the neutron arising from Eq.~(\ref{eq:MnucVol})
are $1/3$ of their value when calculated with PBCs, Eq.~(\ref{deltaM}).~\footnote{If the twist of the up and down 
quarks is the same, $\bm{\phi}^{\pi^{\pm}}$ vanishes and no volume improvement will be obtained by averaging.} 
Of course, calculations at  multiple twist angles need not be performed  to estimate the twist-averaged value, and 
special twist angles can be selected based upon the symmetries of the integer 
sums in Eqs.~(\ref{eq:NNpisum}) and (\ref{eq:NDpisum}).~\footnote{\
This technique is used to reduce the finite-size effects in 
QMC many-body calculations (commonly known as ``special k-points''), as explored in 
Refs.~\cite{PhysRevLett.73.1959, PhysRevB.51.10591, PhysRevB.53.1814, Wilcox:1999ux}.
} 
In particular, it is notable that the leading volume effects of the form 
$e^{-m_\pi L}/L$, $e^{-\sqrt{2}m_\pi L}/L$ and $e^{-\sqrt{3}m_\pi L}/L$, 
can be reduced by a factor of three
with i-PBCs,  by setting the pion twist angle to 
$\bm{\phi}^{\pi^+}=(\frac{\pi}{2},\frac{\pi}{2},\frac{\pi}{2})$.
Averaging  the masses  calculated with PBCs and APBCs also reduces the 
leading contribution by a factor of three.
The leading volume dependence can be eliminated completely by choosing
$\bm{\phi}^{\pi^+}=(\frac{4 \pi}{3},\frac{4\pi}{3},\frac{4\pi}{3})$, leaving  volume corrections to the nucleon mass
of the form $\sim e^{-\sqrt{2} m_\pi {\rm L}}/{\rm L}$.
It is likely that optimal twists exist for other single nucleon properties, such as matrix elements of the isovector axial current, $g_A$.

For arbitrary quark twists, the proton and neutron have, in general,  
different phase spaces as the  momentum modes that exist in the FV differ.
As an example, while quark twists can be chosen to keep the proton at rest in the volume
and allow for averaging over the charged pion twists, 
${\bm\phi}^{(d)} = -2{\bm\phi}^{(u)}$, 
in general the neutron will have non-zero momentum.~\footnote{
Such non-trivial phase spaces somewhat complicate the analysis of LQCD calculations of multi-baryon systems.
}

%%%%%%%%%%%%%%%%%%%%%%%%%%%%%%%%%%%%%%%%%%%%%%%%%%%%%%%%%%
\section{Two Baryons}
\noindent
The positive-energy eigenvalues of two hadrons in a FV subject to PBCs in the spatial directions exhibit 
power-law volume dependences, while the negative-energy eigenvalues deviate exponentially from their 
infinite-volume values.
These energy eigenvalues can be related to the infinite-volume scattering amplitude below the inelastic threshold, 
with corrections that scale as $\sim e^{-m_{\pi}{\rm L}}/{\rm L}$~\cite{Luscher:1986pf,Luscher:1990ux} 
(see also Refs.~\cite{Beane:2003da, Ishizuka:2009bx, Briceno:2013lba, Sato:2007ms}). 
The S-wave NN energy quantization condition (QC)
was generalized to systems with TBCs at rest in Ref.~\cite{Bedaque:2004kc}, 
and to more general two-hadron systems in Ref.~\cite{Agadjanov:2013kja}. 
L\"uscher's energy QC~\cite{Luscher:1986pf, Luscher:1990ux},
which determines the form of the FV corrections,
is dictated by the on-shell two-particle states within the volume.
Once the kinematic constraints on the momentum modes of the two-particle states in the FV are determined, the corresponding 
QC can be determined in a straightforward manner. 
Explicitly, the QC is of the form
\begin{eqnarray}
\det\left[{(\mathcal{M}^{\infty})^{-1}+\delta\mathcal{G}^{V}}\right]\ =\ 0
\ \ \ ,
\label{NNQC}
\end{eqnarray}
where $\mathcal{M}^{\infty}$ is the infinite-volume scattering amplitude matrix evaluated at the on-shell momentum of each particle 
in the CM frame, $p^*$.
For nonrelativistic systems, it is convenient to express the QC in the $\left|JM_J(LS)\right\rangle$ basis, where $J$ is the total angular momentum,  $M_J$ is the eigenvalue of the $\hat
J_z$ operator, and $L$ and $S$ are the orbital angular momentum and the total spin of the system, respectively. 
The matrix elements of $\delta\mathcal{G}^V$ in this basis 
are~\footnote{
This relation has been derived in Ref.~\cite{Briceno:2013bda} for two nucleons  subject to PBCs. 
It reduces to the QC for meson-nucleon scattering~\cite{Bernard:2008ax} and to meson-meson scattering~\cite{Luscher:1990ux, Rummukainen:1995vs, Kim:2005gf, Christ:2005gi}.
For an alternate derivation, see Ref.~\cite{Ishizuka:2009bx}. 
}
\begin{eqnarray}
&& \left[\delta\mathcal{G}^V\right]_{JM_J,LS;J'M_J',L'S'}=i	\eta \frac{p^*}{8\pi E^*}
\delta_{SS'}\left[\delta_{JJ'}\delta_{M_JM_J'}\delta_{LL'} +i\sum_{l,m}\frac{(4\pi)^{3/2}}{p^{*l+1}}
c_{lm}^{\mathbf{d},\bm{\phi}_1,\bm{\phi}_2}(p^{*2};{\rm L}) \right.
\nonumber\\
&& \qquad \qquad \qquad \left .  \times \sum_{M_L,M_L',M_S}\langle JM_J|LM_L,SM_S\rangle \langle L'M_L',SM_S|J'M_J'\rangle 
\int d\Omega~Y^*_{L M_L}Y^*_{l m}Y_{L' M_L'}\right],
\label{deltaG}
\end{eqnarray}
where $\eta=1/2$ for identical particles and $\eta=1$ otherwise,
and $\langle JM_J|LM_L,SM_S\rangle$ are Clebsch-Gordan coefficients. 
$E^*$ is the total CM energy of the system, $E^*=\sqrt{p^{*2}+m_1^2}+\sqrt{p^{*2}+m_2^2}$ where
$m_1$ and $m_2$ are the masses of the particles, 
and $\bm{\phi}_1$  and $\bm{\phi}_2$ are their respective twist angles. 
The total momentum of the system is  $\mathbf{P}=\frac{2\pi}{{\rm L}}\mathbf{d}+\frac{\bm{\phi}_1+\bm{\phi}_2}{{\rm L}}$ 
with $\mathbf{d}\in \mathbb{Z}^3$. 
 The volume dependence and the dependence on the BCs are in the kinematic functions 
 $c_{lm}^{\mathbf{d},\bm{\phi}_1,\bm{\phi}_2}(p^{*2};{\rm L})$, 
 defined as
 \begin{eqnarray}
c^{\textbf{d},\bm{\phi}_1,\bm{\phi}_2}_{lm}(p^{*2};{\rm L})
\ =\ \frac{\sqrt{4\pi}}{\gamma {\rm L}^3}\left(\frac{2\pi}{{\rm L}}\right)^{l-2}
\mathcal{Z}^{\mathbf{d},\bm{\phi}_1,\bm{\phi}_2}_{lm}[1;(p^*{\rm L}/2\pi)^2]
\ \ \ ,
\label{clm}
\end{eqnarray}
with
\begin{eqnarray}
\mathcal{Z}^{\mathbf{d},\bm{\phi}_1,\bm{\phi}_2}_{lm}[s;x^2]
\ =\ \sum_{\mathbf r \in \mathcal{P}_{\mathbf{d},\bm{\phi}_1,\bm{\phi}_2}}
\frac{ |{\bf r}|^l \ Y_{l m}(\mathbf{r})}{(\mathbf{r}^2-x^2)^s}
\ \ \ .
\label{Zlm}
\end{eqnarray}
$\gamma=E/E^*$ where E is the total energy of the system in the rest frame of the volume (the lab frame), 
$E^2=\mathbf{P}^2+E^{*2}$. 
The sum in Eq.~(\ref{Zlm}) is performed over the momentum vectors $\mathbf{r}$ that belong to the 
set $\mathcal{P}_{\mathbf{d},\bm{\phi}_1,\bm{\phi}_2}$, 
which remains to be determined.  
 
Consider the two-hadron wavefunction in the lab frame~\cite{Rummukainen:1995vs, Davoudi:2011md} 
that is subject to the TBCs,
\begin{eqnarray}
\psi_{{\rm Lab}}(\mathbf{x}_1+{\rm{L}}\mathbf{n}_1,\mathbf{x}_2+{\rm{L}}\mathbf{n}_2)
\ =\ 
e^{i{\bm{\phi}}_1 \cdot \mathbf{n}_1+i{\bm{\phi}}_2 \cdot \mathbf{n}_2}
\ \psi_{{\rm Lab}}(\mathbf{x}_1,\mathbf{x}_2)
\ \ \ ,
\label{WF-BC}
\end{eqnarray}
where $\mathbf{x}_1$ and $\mathbf{x}_2$ denote the position of the hadrons, 
and $\mathbf{n}_1,\mathbf{n}_2\in\mathbb{Z}^3$. 
As the total momentum of the system is conserved, the wavefunction can be written as an eigenfunction of 
the total momentum $P=(E,\mathbf{P})$. 
In the lab frame, the equal-time wavefunction of the system is
\begin{eqnarray}
\psi_{{\rm Lab}}(x_1,x_2) 
\ =\  e^{-iE X^0+i\mathbf{P} \cdot \mathbf{X}}
\ \varphi_{{\rm Lab}}(0,\mathbf{x}_1-\mathbf{x}_2)
\ \ \ ,
\end{eqnarray}
where the position of the CM is $X$, and
\begin{eqnarray}
X & = & \alpha x_1+(1-\alpha) x_2
\ \ \ \ , \ \ \ \ 
\alpha=\frac{1}{2}\left(1+\frac{m_1^2-m_2^2}{E^{*2}}\right)
\ \ \ ,
\end{eqnarray}
for  systems with unequal masses~\cite{Davoudi:2011md}.  
Since the CM wavefunction is independent of the relative time coordinate~\cite{Rummukainen:1995vs},  
$\varphi_{{\rm Lab}}(0,\mathbf{\mathbf{x}_1-\mathbf{x}_2})=\varphi_{{\rm CM}}(\hat{\gamma} (\mathbf{x}_1-\mathbf{x}_2))$,
where the boosted relative position vector is
$\hat{\gamma} \mathbf{x}=\gamma \mathbf{x}_{\Vert}+ \mathbf{x}_{\bot}$, 
with $\mathbf{x}_{\Vert}$ ($\mathbf{x}_{\bot}$) the component of $\mathbf{x}$ that is 
parallel (perpendicular) to $\mathbf{P}$. 
By expressing $\psi_{{\rm Lab}}$ in  Eq.~(\ref{WF-BC}) in terms of $\varphi_{CM}$, it straightforwardly follows that
\begin{eqnarray}
e^{i\alpha\mathbf{P}\cdot(\mathbf{n}_1-\mathbf{n}_2){\rm L}+i\mathbf{P}\cdot\mathbf{n}_2{\rm L}}
\ \varphi_{{\rm CM}}(\mathbf{y}^*+\hat{\gamma}(\mathbf{n}_1-\mathbf{n}_2){\rm L})
\ =\ 
e^{i{\bm{\phi}}_1 \cdot \mathbf{n}_1+i{\bm{\phi}}_2 \cdot \mathbf{n}_2}
\ \varphi_{{\rm CM}}(\mathbf{y}^*)
\ \ \ \ ,
\end{eqnarray}
where $\mathbf{y}^*=\mathbf{x}_1^*-\mathbf{x}_2^*$ is the relative coordinate of two hadrons in the CM frame. 
By Fourier transforming this relation, and using the form of the total momentum $\mathbf{P}$ from above, 
the relative momenta allowed in the FV energy QC are constrained to be 
\begin{eqnarray}
\mathbf{r}
\ =\ 
\frac{1}{{\rm L}}\ 
\hat{\gamma}^{-1}
\ \left[2\pi(\mathbf{n}-\alpha\mathbf{d})-(\alpha-\frac{1}{2})(\bm{\phi}_1+\bm{\phi}_2)+\frac{1}{2}(\bm{\phi}_1-\bm{\phi}_2)\right]
\ \ \ \ ,
\label{r-TBC}
\end{eqnarray}
where $\mathbf{n}\in\mathbb{Z}^3$ is  summed over in Eq.~(\ref{Zlm}).
These results encapsulate those of Refs.~\cite{Rummukainen:1995vs, Davoudi:2011md, Fu:2011xz, Leskovec:2012gb, Bour:2011ef} 
when the PBCs are imposed, 
i.e.,  when $\bm{\phi}_1=\bm{\phi}_2=\mathbf{0}$. 
It also recovers two limiting cases that are considered in Ref.~\cite{Agadjanov:2013kja} for the use of TBCs in the scalar sector of QCD. 
It should be noted that for particles with equal masses, $\alpha=1/2$, 
the set of allowed momentum vectors reduces to
\begin{eqnarray}
\mathbf{r}
\ =\ 
\frac{1}{{\rm L}}\ \hat{\gamma}^{-1}
\ \left[2\pi(\mathbf{n}-\frac{1}{2}\mathbf{d})+\frac{1}{2}(\bm{\phi}_1-\bm{\phi}_2)\right]
\ \ \ .
\label{eq-mass}
\end{eqnarray}
It is important to note that for two identical hadrons, when $\bm{\phi}_1=\bm{\phi}_2=\bm{\phi}$, the FV spectra show no non-trivial 
dependence on the twist other than a  shift in the total energy of the system, 
$E^2=(\frac{2\pi}{{\rm L}}\mathbf{d}+\frac{\bm{\phi}}{{\rm L}})^2+E^{*2}$. 
As a result, twisting will not provide additional constraints on the scattering amplitude in, 
for instance, the $\si$ nn or pp channels.
This is also the case for the FV studies of NN scattering in the $\siii$-$\diii$ coupled channels if the same twist is 
imposed on the up and down quarks.~\footnote{This result differs somewhat from the conclusion of Ref.~\cite{Bedaque:2004kc}.}

%%%%%%%%%%%%%%%%%%%%%%%%%%%%%%%%%%%%%%%%%%%%%%%%%%%%%%%%%%
\subsection{The Deuteron}
\noindent
The energy spectra of two nucleons  with spin $S=1$ in a FV subject to PBCs 
and with a range of CM momenta have been determined from the 
experimentally measured phase shifts and mixing angles~\cite{Briceno:2013bda}.
In particular,  the dependence of the bound-state spectra on the non-zero mixing angle between S and D waves, $\epsilon_1$, 
has been determined. 
As seen from Eq.~(\ref{eq-mass}), the 
effects of the 
twist angles $\frac{1}{2\pi}(\bm{\phi}_1~-~\bm{\phi}_2)=(0,0,1),(1,1,0),(1,1,1)$ 
on the CM spectra
are the same as those of  (untwisted) boost vectors, ${\bf d}$,
considered in Ref.~\cite{Briceno:2013bda}.
Therefore,  different TBCs can provide additional CM energies in a single volume, 
similar to boosted calculations, 
which can be used to better constrain  scattering parameters and the S-matrix. 
However, twisting may be a more powerful tool as it 
provides access to a continuum of momenta.

If imposing TBCs on the quark fields would require the generation of new ensembles of gauge-field configurations,
it would likely not be optimal to expend  large computational resources  on multiple twisted calculations.
However,  PBCs can be retained on the sea quarks  and 
TBCs can be imposed only in the valence sector~\cite{Bedaque:2004ax}. 
The reason for this is that there are no disconnected diagrams  associated with the NN interactions.~\footnote{
As  recently demonstrated, disconnected diagrams will not hinder the use of partially-TBCs in studies of the scalar 
sector of QCD either~\cite{Agadjanov:2013kja}. 
The graded symmetry of ``partially-quenched'' QCD  results in  cancellations among 
contributions from intermediate non-valence mesons.} 
At the level of the low-energy EFT, this indicates that there are no intermediate s-channel diagrams in which a nucleon or meson 
containing a sea quark can go on-shell. 
Such off-mass-shell hadrons modify the NN interactions by  $\sim e^{-m_{\pi}{\rm L}}/{\rm L}$, and do not invalidate the use 
of the QC in  Eq.~(\ref{NNQC}) with the partially-TBCs as long as the calculations are performed
in sufficiently large volumes,  
${\rm L} \gtrsim 9 ~ {\rm fm}$.

One significant advantage of imposing TBCs is the improvement in the volume dependence of the deuteron binding energy. 
Although  the formalism presented in the previous section can be used to fit to various scattering parameters~\cite{Briceno:2013bda} 
(and consequently determine the deuteron binding energy), 
we will show that with a judicious choice of twist angles, the extracted energies in future LQCD calculations should be close to 
the infinite-volume values,  even in volumes as small as $\sim (9~{\rm fm})^3$.

As discussed in the previous section, the CM energy of the np system is sensitive to TBCs only if 
different twists are imposed upon the up and down quarks. 
This means that, even if exact isospin symmetry is assumed, the proton and the neutron will have different phase spaces
due to the different BCs. 
By relaxing the interchangeability constraint on the np state, as required by the different phase spaces, 
the NN positive-parity channels will mix with the negative-parity channels. 
This admixture of parity eigenstates is entirely a FV effect induced by the boundary conditions, 
and does not require parity violation in the interactions, 
manifesting itself in  non-vanishing $c_{lm}^{\mathbf{d},\bm{\phi}_1,\bm{\phi}_2}$ functions 
for odd values of $l$.  
As such, the spin of the NN system is preserved.

The QC in Eq.~(\ref{NNQC}) depends on S-matrix elements in all partial waves, however it can be truncated
to include only channels with $L \leq 2$
(requiring $J \leq 3$) 
because of the reducing size of the  low-energy phase shifts in the higher channels. 
For arbitrary twist angles, the truncated QC can be represented by a $27 \times 27$ matrix in the $\left|JM_J(LS)\right\rangle$ basis, the eigenvalues of which dictate the energy eigenvalues.
Fits to the experimentally known phase shifts and 
mixing parameters~\cite{Nijmegen, PhysRevC.48.792, PhysRevC.49.2950, PhysRevC.54.2851, PhysRevC.54.2869} 
are used to extrapolate to negative 
energies~\cite{Briceno:2013bda} to provide the inputs into the truncated QC,
from which the deuteron spectra in a cubic volume with TBCs are predicted. 
The scattering parameters entering the  analysis are 
$\delta_{1\alpha},~\epsilon_1,~\delta_{1\beta},~\delta^{({^3}P_0)},~\delta^{({^3}P_1)},~\delta^{({^3}P_2)},
~\delta^{({^3}D_2)}$ and $\delta^{({^3}D_3)}$, 
where the Blatt-Biedenharn (BB) parameterization~\cite{Blatt:1952zza} is used in the $J=1$ sector. 
The twist angles explored in this work are 
$\bm{\phi}^p=-\bm{\phi}^n \equiv \bm{\phi}=(0,0,0)$ (PBCs), 
$(\pi,\pi,\pi)$ (APBCs)
and $(\frac{\pi}{2},\frac{\pi}{2},\frac{\pi}{2})$ (i-PBCs).
 At the level of the quarks, this implies that the twist angles of the (valence) up and down quarks 
are ${\bm\phi}^u =  -{\bm\phi}^d = {\bm\phi}$.
We also set $\mathbf{d}=\mathbf{0}$ in Eq.~(\ref{r-TBC}) so that the np system is at rest in the lab frame. 
The reason for this choice of twist angles is  that they (directly or indirectly) give rise a significant cancellation of the 
leading FV corrections to the masses of the nucleons, as shown in Sec.~\ref{sec:Single}. 
The number of eigenvalues of 
${(\mathcal{M}^{\infty})^{-1}+\delta\mathcal{G}^{V}}$, 
and their degeneracies, reflect the spatial-symmetry group of the FV. 
Calculations with $\bm{\phi}=\bm{0}$ respect the cubic ($O_h$) symmetry, while 
for $\bm{\phi}=(\frac{\pi}{2},\frac{\pi}{2},\frac{\pi}{2})$ the symmetry group is reduced to the $C_{3v}$ point 
group.~\footnote{
There is a correspondence between the FV spatial symmetry in twisted calculations with  
arbitrary twists
$\bm{\phi}^p \neq \bm{\phi}^n$ 
and the FV symmetry in (boosted) NN calculations with PBCs when isospin breaking is considered. 
For example, the point symmetry group corresponding to twisted calculations with $\bm{\phi}^p = -\bm{\phi}^n=(0,0,\frac{\pi}{2})$ 
and that of the physical np system with $\mathbf{P}=\frac{2\pi}{\rm L}(0,0,1)$ with PBCs are both $C_{4v}$. 
} 
However, for $\bm{\phi}=(\pi,\pi,\pi)$ 
the system has inversion symmetry, and respects the $D_{3h}$ point symmetry~\cite{Dresselhaus}. 
By examining the transformation properties of the 
$c_{lm}^{\mathbf{d},\bm{\phi}_1,\bm{\phi}_2}$ 
functions under the symmetry operations of these groups, certain relations are found for any given $l$. 
These relations, 
as well as the eigenvectors of the FV matrices,
which are tabulated elsewhere~\cite{Luscher:1990ux, Rummukainen:1995vs, Briceno:2013lba, Gockeler:2012yj,Thomas:2011rh,Dudek:2012gj},
can be used to block diagonalize the $27 \times 27$ matrix representation of the  QCs, 
where each block corresponds to an irrep of the point-group symmetry of the system. 
For the selected twist angles, the QCs of the irreps of the corresponding point 
groups that have overlap with the deuteron are given in the Appendix~\ref{app: QC} .

\begin{figure}[h]
\begin{center}
\includegraphics[scale=0.40]{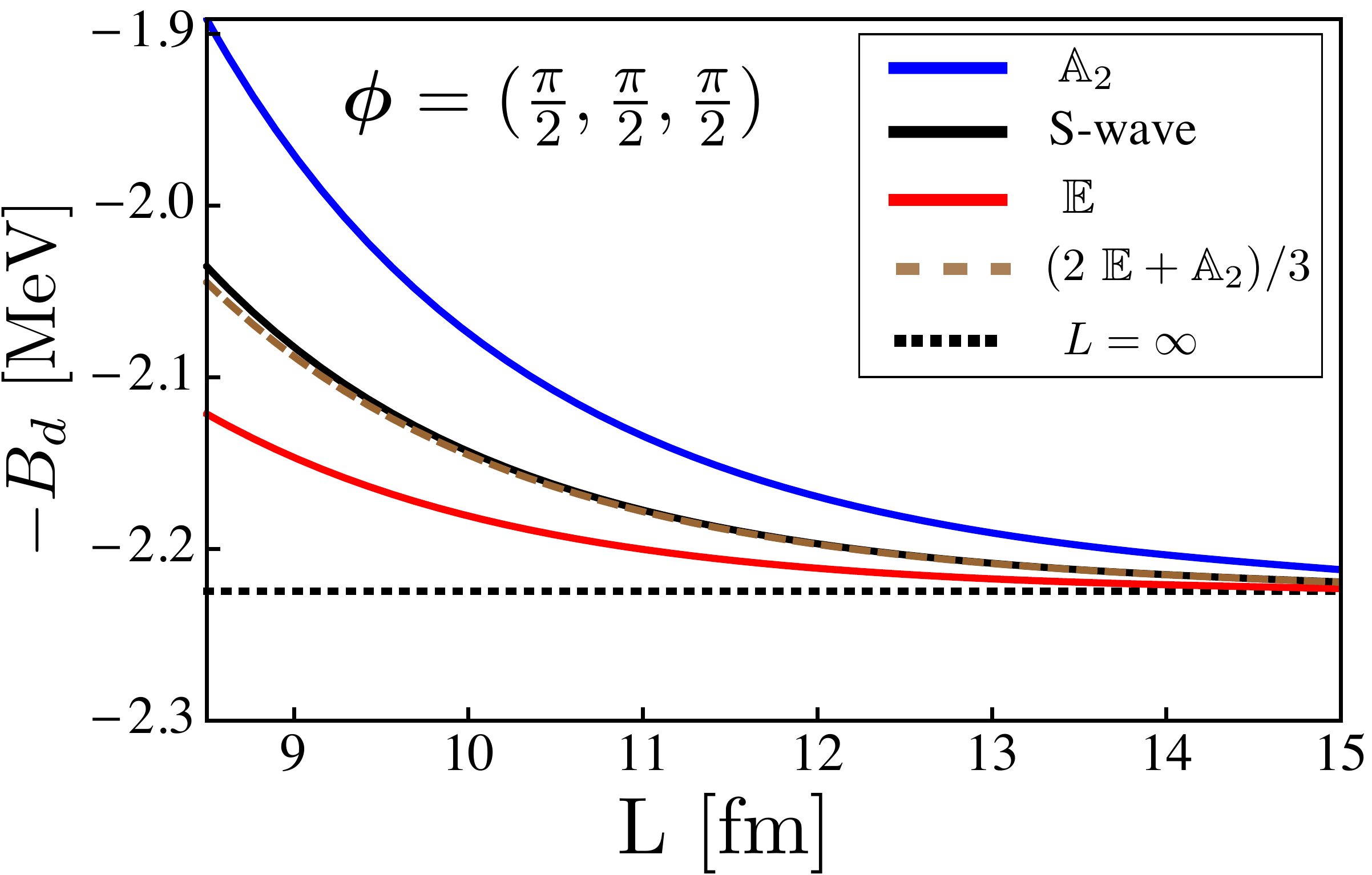}
\caption{
The deuteron binding energy as a function of ${\rm L}$ 
using i-PBCs
($\bm{\phi}^p=-\bm{\phi}^n \equiv \bm{\phi}=(\frac{\pi}{2},\frac{\pi}{2},\frac{\pi}{2})$). 
The blue curve corresponds to the $\mathbb{A}_2$ irrep of the $C_{3v}$ group, 
while the red curve corresponds to the $\mathbb{E}$ irrep. 
The brown-dashed curve corresponds to the weighted average of the $\mathbb{A}_2$ and 
$\mathbb{E}$ irreps, $-\frac{1}{3}(2B_d^{(\mathbb{E})}+B_d^{(\mathbb{A}_2)})$,
while the black-solid curve corresponds to the S-wave limit.
The infinite-volume deuteron binding energy is shown  by the black-dotted line.
}
\label{fig:A2-E}
\end{center}
\end{figure}
\begin{figure}[h]
\begin{center}
\subfigure[]{
\includegraphics[scale=0.33]{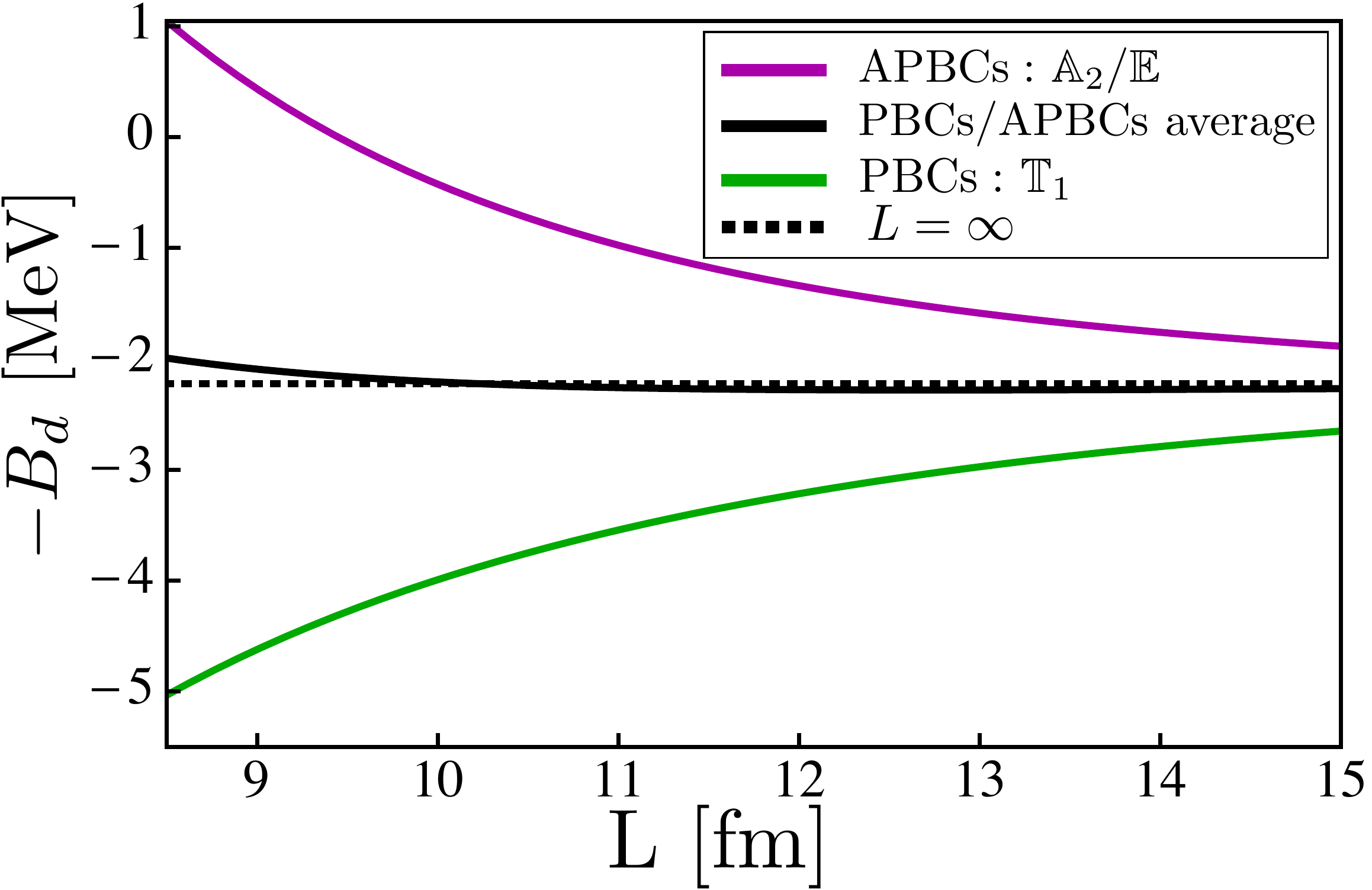}}
\subfigure[]{
\includegraphics[scale=0.33]{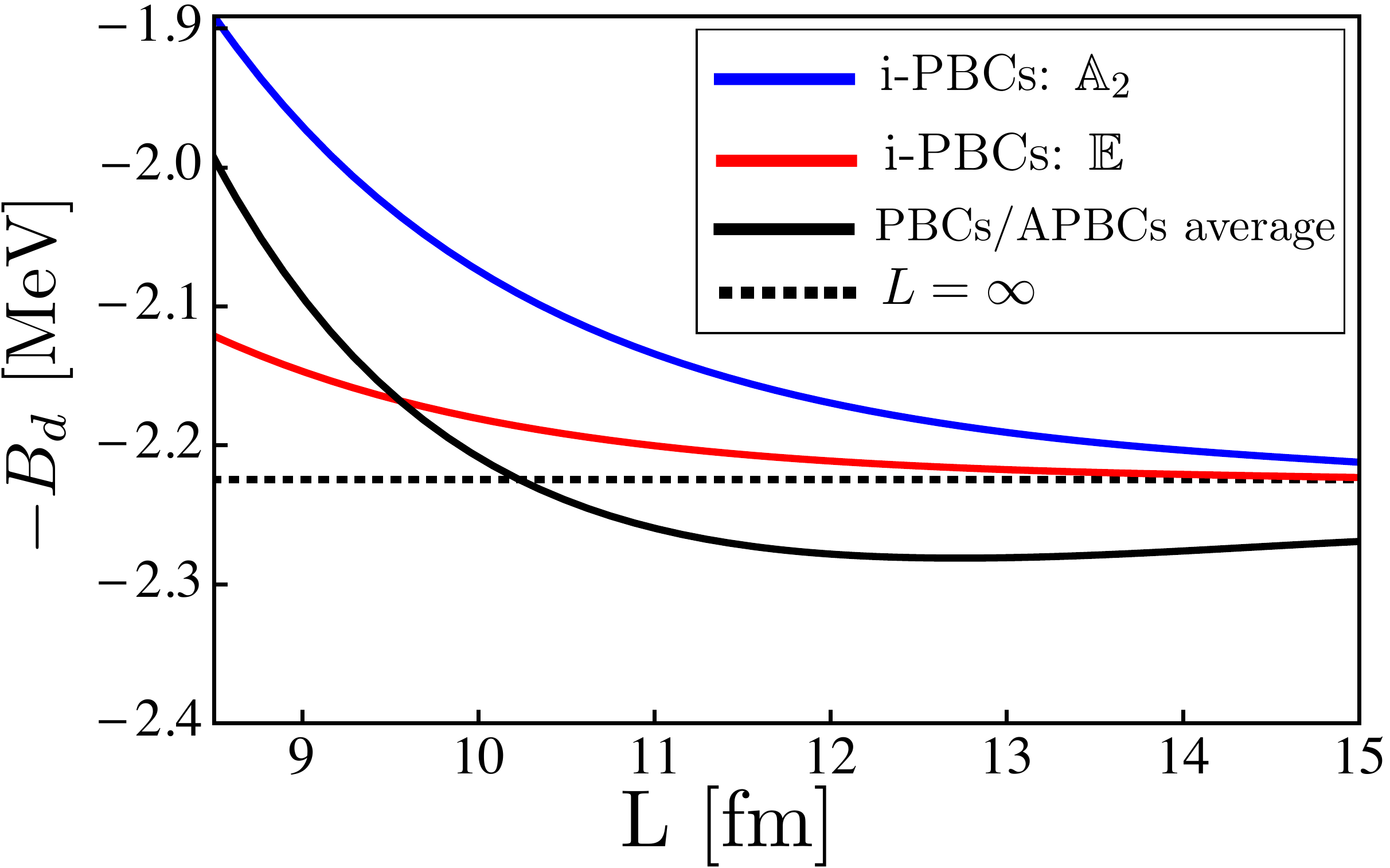}}
\caption{
a) The deuteron binding energy as a function of ${\rm L}$ 
from  PBCs (green curve) and  from APBCs (purple curve). 
The black-solid curve represents the average of these energies. 
b) A closer look at the average in part (a) compared with  energies obtained with i-PBCs,
$\mathbb{A}_2$ (blue curve) and $\mathbb{E}$ (red curve).
}
\label{fig:PBC-APBC}
\end{center}
\end{figure}

For i-PBCs, there are two irreps of the $C_{3v}$ group, namely the one-dimensional irrep $\mathbb{A}_1$ 
and the two-dimensional irrep $\mathbb{E}$, that have overlap with the $^3{S_1}$-${^3}{D_1}$ coupled channels. 
Fig.~\ref{fig:A2-E} shows the binding energy (the CM energy minus the rest masses of the nucleons), $-B_d=E^*-M_p-M_n$, 
as a function of ${\rm L}$ corresponding to $\mathbb{A}_2$ irrep (blue curve) and $\mathbb{E}$ irrep (red curve), 
obtained from the QCs in Eqs. (\ref{pi2pi2pi2A2}) and (\ref{pi2pi2pi2E}).  
Even at ${\rm L}\sim9~{\rm fm}$, the deuteron binding energies extracted from both irreps are close to the infinite-volume value. 
In particular, calculations in the $\mathbb{E}$ irrep of the $C_{3v}$ group provide a few percent-level accurate 
determination of the deuteron binding energy 
in this volume. 
The black-solid curve in Fig.~\ref{fig:A2-E} 
represents the S-wave limit of the interactions, 
when the S-D mixing parameter and all phase shifts except that in the S-wave are set equal to zero. 
The $M_J^\prime$-averaged binding energy, 
$-\frac{1}{3}(2B_d^{(\mathbb{E})}+B_d^{(\mathbb{A}_2)})$, 
converges to this S-wave limit, as shown 
in Fig. \ref{fig:A2-E}
(the $\mathbb{A}_2$ irrep contains the $M_J^\prime=0$ state 
while $\mathbb{E}$ contains the $M_J^\prime=\pm 1$ states,
where $M_J^\prime$ is the projection of total angular momentum
along the twist direction). 
In order to appreciate the significance of calculations performed with the $\bm{\phi}=(\frac{\pi}{2},\frac{\pi}{2},\frac{\pi}{2})$ twist angles, 
it is helpful to recall the deuteron binding energy obtained in calculations with PBCs. 
For PBCs, 
the only irrep of the cubic group that has overlap with the $\siii$-$\diii$ coupled channels is the 
three-dimensional irrep $\mathbb{T}_1$, Eq.~(\ref{000T1}), 
and the corresponding binding energy is shown in Fig. \ref{fig:PBC-APBC}(a) (green curve). 
As is well known, the binding energy deviates significantly from its infinite-volume value, 
such that the FV deuteron is approximately twice as bound as the infinite-volume deuteron at ${\rm L}=9~{\rm fm}$. 
For APBCs, two irreps of the $D_{3h}$ group overlap with the deuteron channel, 
$\mathbb{A}_2$ and $\mathbb{E}$ (Eqs. (\ref{pipipiA2},\ref{pipipiE})), 
and yield degenerate binding energies as shown in Fig. \ref{fig:PBC-APBC}(a) (purple curves). 
As seen in Fig. \ref{fig:PBC-APBC}(a), 
the deuteron becomes unbound over a range of volumes 
and  asymptotes slowly to the infinite-volume limit. 
However, in analogy with the nucleon masses, 
the volume dependence of the deuteron binding energy 
is significantly reduced 
by averaging the results obtained with PBCs and APBCs,
as shown in Fig. \ref{fig:PBC-APBC}(a) (black-solid curve). 
Fig. \ref{fig:PBC-APBC}(b) provides a magnified  view of this averaged quantity (black-solid curve),
where the two energy levels associated  with i-PBCs are shown for comparison.

In order to understand the observed volume improvements,  
consider the volume scaling of the full QC assuming that the phase shifts beyond the $\alpha$-wave 
are small.
In this limit, for a general set of twist angles and boosts, the QC  collapses to 
\begin{eqnarray}
&&
\det\left[
\left(
p^*\cot\delta_{1\alpha} - 4\pi c_{00} 
\right)
\left(
\begin{array}{ccc}
1&0&0\\ 0&1&0\\ 0&0&1
\end{array}
\right)
\right.\nonumber\\
&&
\qquad\qquad
\left.
- 
{2\pi\over \sqrt{5} p^{*2}}\left(\sqrt{2}\sin 2\epsilon_1 - \sin^2\epsilon_1 \right)
\left(
\begin{array}{ccc}
c_{20}&\sqrt{3} c_{21}&\sqrt{6} c_{22}\\
-\sqrt{3} c_{2-1}&-2c_{20}&-\sqrt{3} c_{21}\\
\sqrt{6} c_{2-2}&\sqrt{3} c_{2-1}&c_{20}
\end{array}
\right)
\ \right]  \ = \  0
\ \ \ ,
\label{eq:QC3by3}
\end{eqnarray}
which depends upon the $\alpha$-wave phase shift and the mixing parameter, $\epsilon_1$.
Shorthand notation has been used for convenience, $c_{lm} = c_{lm}^{{\bf d},{\bm\phi}_1,{\bm\phi}_2}(p^{*2}; {\rm L})$.
For generic twist angles,
deviations between the energy eigenvalues resulting from this truncated QC and the full QC scale as 
$\sim \tan\delta_i \ e^{-2 \kappa {\rm L}}/(\kappa {\rm L}^2)$,
where $\delta_i$ denotes phase shifts beyond the $\alpha$-wave
(see Appendix~\ref{app:TwistC} for  expansions of the $c_{lm}^{{\bf d},{\bm\phi}_1,{\bm\phi}_2}$ functions).
For i-PBCs, the leading corrections are from the P-waves, as can be seen from the expansions of the $c_{lm}$ in Table~\ref{app:TwistC}.
By neglecting the small mixing between the S-wave and D-waves in Eq.~(\ref{eq:QC3by3}), 
the QC  dictated by  S-wave interactions is~\footnote{
In the limit where $\epsilon_1=0$, the $J=1$ $\alpha$-wave is entirely S-wave, 
while the $\beta$-wave is entirely D-wave. 
This approximation neglects FV effects of the form $\epsilon_1 e^{- \kappa {\rm L}}/{\rm L}$.
}
\begin{eqnarray}
p^*\cot\delta^{{(^3S_1)}}|_{p^*=i\kappa}+\kappa=\sum_{\mathbf{n}\neq\mathbf{0}}
e^{i  (\alpha-\frac{1}{2}) \mathbf{n} \cdot (\bm{\phi}^p+\bm{\phi}^n)} e^{-i \frac{1}{2} \mathbf{n} \cdot (\bm{\phi}^p-\bm{\phi}^n)}
e^{i 2\pi \alpha \mathbf{n} \cdot \mathbf{d}}
~\frac{e^{-|\hat{\gamma}\mathbf{n}| \kappa {\rm L}}}{|\hat{\gamma}\mathbf{n}|{\rm L}}
\ \ \ .
\label{S-QC}
\end{eqnarray}
The volume dependence of the deuteron binding momentum, $\kappa$, originates from the right-hand side of this equation.
For ${\bf d}={\bf 0}$,  the $c_{2m}$ functions vanish  for both PBCs and APBCs, leading to Eq.~(\ref{S-QC}) without further approximation.
For the twist angles $\bm{\phi}^p=-\bm{\phi}^n \equiv \bm{\phi}=(\frac{\pi}{2},\frac{\pi}{2},\frac{\pi}{2})$ and boost $\mathbf{d}=\mathbf{0}$,
the first few terms in the summation on the right-hand side of Eq.~(\ref{S-QC}) ($\mathbf{n}^2\leq 3$)
vanish, leaving the leading volume corrections to scale as $\sim e^{-2\kappa{\rm L}}/{\rm L}$.
A lesser 
cancellation occurs in the average of binding energies obtained with PBCs and APBCs, giving rise to deviations from the infinite-volume energy by terms that scale as $\sim e^{-\sqrt{2}\kappa{\rm L}}/{\rm L}$.

The result of Monte Carlo twist averaging of the deuteron binding energy 
can be ascertained from the behavior of the two extreme contributions, the PBC and APBC results.
While the average binding energy obtained from $N$ randomly selected sets of twist angles scales as 
$B_d^{(\infty)} +  {\cal O}\left( e^{-2\kappa{\rm L}}/{\rm L} \right)$, 
the standard deviation of the mean scales as $\sim e^{-\kappa{\rm L}}/(\sqrt{N} {\rm L})$, 
giving rise to a signal-to-noise in the binding energy that scales as 
$\sim  \sqrt{N} \ B_d^{(\infty)}\  {\rm L}\ e^{\kappa{\rm L}}$, which even for ${\rm L}\sim 14~{\rm fm}$ allows only for a poor extraction, as 
can be deduced from Fig. \ref{fig:PBC-APBC}(a).
It is clear that such a method is inferior to that of pair-wise averaging, such as from PBCs and APBCs, or choosing special twists, such as 
i-PBCs.

We have restricted ourselves to the scenarios where the net twist angles in each 
Cartesian direction (the lattice axes) are the same. 
One reason for this is that systems with arbitrary twists give rise to three distinct, but nearby, energy eigenvalues 
associated with combinations of each of the three $M_J$-states of the deuteron - a sub-optimal system to analyze
in LQCD calculations.
Another reason is that a twist of $\frac{\pi}{2}$  in each direction is optimal in minimizing the FV effects in 
both the two-body binding energies and the single-baryon masses.
Further, averaging the results of calculations with PBCs and APBCs also eliminates the leading FV corrections to both quantities.
We re-emphasize that ultimately, one wants to extract as many scattering parameters as feasible from calculations 
in a single volume, requiring calculations with multiple boosts of the CM as well as multiple arbitrary twists, in order  
to maximize the inputs to the energy QCs. 
In general, with arbitrary twist angles,  $\bm{\phi}=(\phi_x,\phi_y,\phi_z)$,
the $27 \times 27$ matrix representation of the QC matrix cannot be block diagonalized and it has $27$ distinct eigenvalues.
The truncation to the $3\times 3$ matrices given in Eq.~(\ref{eq:QC3by3}) remains valid, as do the estimates of the truncation errors, but this truncated QC will provide three distinct energy eigenvalues.

While not the focus of this work, it is worth reminding ourselves about the behavior of the positive-energy states in the FV,
such as the higher states associated with the $\siii$-$\diii$ coupled channel or those associated with the $\si$ $np$ channel,
as described in Eq.~(\ref{NNQC}).
For an arbitrary twist, the non-interacting energy levels in the FV are determined by integer triplets and the twist angles.
Interactions will produce deviations from these non-interacting levels, that become smaller
as the lattice volume increases, scaling with $\sim \tan\delta(p^*)/(M {\rm L}^2)$.
As discussed previously, as there is no underlying symmetry for arbitrary twists, the eigenstates will, in general, be non degenerate.

%%%%%%%%%%%%%%%%%%%%%%%%%%%%%%%%%%%%%%%%%%%%%%%%%%%%%%%%%%
\section{Summary and Conclusions}
\noindent
Twisted boundary conditions have been successfully used in numerical calculations of important observables, 
both in nuclear and particle physics with Lattice QCD, 
as well as in others areas such as condensed-matter physics.
They provide a means with which to select the phase space of particles  in a given finite volume, beyond that allowed by periodic or anti-periodic boundary conditions.
In LQCD calculations, TBCs have been used to resolve the threshold region required in the evaluation of 
transition matrix elements without requiring large lattice volumes \cite{Tiburzi:2005hg, Jiang:2006gna, Boyle:2007wg, Simula:2007fa, Boyle:2008yd, Aoki:2008gv, Boyle:2012nb, Brandt:2013mb,Boyle:2013gsa}. 
They can also be
used in calculations of elastic $2 \rightarrow 2$ processes by providing a better sampling of 
CM kinematics in a single volume, allowing for 
better constraints on scattering parameters~\cite{Bernard:2010fp, Doring:2011vk, Doring:2012eu, Ozaki:2012ce}. 
In this paper, we have explored the use of TBCs in calculating the mass of single baryons, and in determining the binding of two-hadron 
systems in a FV, with a focus on the deuteron.  
In particular, we have used experimentally known scattering data to determine the location of the lowest-lying 
FV states that have overlap with the deuteron for a selection of twist angles, and combinations thereof.
We have found that twisting provides an effective way of exponentially reducing the impact of the finite lattice volume on the calculation of 
two-body binding energies.  Pair-wise combining results obtained with particular twists, such as 
PBCs and APBCs, 
can eliminate the leading volume dependence.  
The same is true for twist averaging, but the uncertainty resulting from a finite number of randomly selected twists can be large.
Importantly, we have 
determined that the i-PBCs, with 
$\bm{\phi}=(\frac{\pi}{2},\frac{\pi}{2},\frac{\pi}{2})$,
eliminate the first three FV corrections to the dominant S-wave contribution to 
the two-hadron binding energies, suppressing such effects from 
$\mathcal{O}\left(e^{-\kappa {\rm L}}/{\rm L}\right)$
to
$\mathcal{O}\left(e^{- 2 \kappa{\rm L}}/{\rm L}\right)$, while also reducing the FV modifications to the nucleon mass, of the form
$\mathcal{O}\left(e^{- m_\pi {\rm L}}/{\rm L}\right)$, by a factor of three.
This translates into at least an order of magnitude improvement in the accuracy of the deuteron binding energy 
extracted from  LQCD correlation functions in volumes as small as $\sim (9~{\rm fm})^3$. 
As partially-TBCs modify the nuclear forces by terms of order $\mathcal{O}\left(e^{-m_{\pi}{\rm L}}/{\rm L}\right)$,
such calculations of the deuteron and other bound states can be performed without the need for 
multiple ensembles of gauge-field configurations, significantly reducing the required computational resources.

Given the generalized L\"uscher FV formalism for NN systems \cite{Briceno:2013lba} with TBCs, 
not only can the binding energy of the deuteron be obtained from the upcoming LQCD calculations, 
but the relevant scattering parameters, including the S-D mixing parameter, can be well constrained. 
While giving different  twists to the up and down quarks modifies the neutron and proton phase 
space in different ways that allows for a parametric reduction in volume effects to the deuteron binding energy, 
and control on the location of the positive-energy scattering states,
it does not change the CM phase space in the neutron-neutron or proton-proton systems.  
Therefore, it is  not a useful tool in 
refining calculations of scattering parameters in these channels.

Inspired by the volume improvement seen in the QMC calculations of few and many-body systems with 
twist-averaged BCs~\cite{2001PhRvE..64a6702L, PhysRevLett.73.1959, PhysRevB.51.10591, PhysRevB.53.1814, Wilcox:1999ux}, 
and  studies of Dirichlet BCs and PBCs in QMC and Density-Functional Theory, e.g. Refs.~\cite{Bulgac:2013mz,Erler:2012qd},
and considering the twist-phase modifications to the images associated with a given system,
we speculate that the FV modifications to the spectrum of three-nucleon and multi-nucleon systems can  be   reduced 
by TBCs.  The magnitude of the improvement  
will depend upon the inter-particle forces being short ranged compared to the extent of the system. 
Due to the complexity of such systems, particularly in a FV~\cite{Polejaeva:2012ut, Briceno:2012rv,Hansen:2013dla}, a definitive 
conclusion can only be arrived at upon further investigation.

%%%%%%%%%%%%%%%%%%%%%%%%%%%%%%%%%%%%%%%%%%%

\noindent
\subsection*{Acknowledgments}
\noindent
We would like to thank W. Nazarewicz, S. Reddy,
and other attendees of the INT program 
{\it Quantitative Large Amplitude Shape Dynamics: Fission and Heavy-Ion Fusion}
(September-November 2013), for stimulating discussions.
RB acknowledges support from the US DOE contract DE-AC05-06OR23177, under which 
Jefferson Science Associates, LLC, manages and operates the Jefferson Laboratory.
The work of TL was supported by the DFG through SFB/TR 16 and SFG 634.
ZD and MJS were supported in part by DOE grant No. DE-FG02-00ER41132. 

%%%%%%%%%%%%%%%%%%%%%%%%%%%%%%%%%%%%%%
\bibliography{bibi}

\newpage

\appendix
%\begin{small}
\section{Twisted Images 
\label{app: TI}
}
%%%%%%%%%%%%%%%%%%%%%%%%%%%%%%%%%%%%%%
\noindent
It is helpful to make explicit the sums over the twist phases. 
Consider the sum  
\begin{eqnarray}
S({\bm\phi}) & = & 
\sum_{\mathbf{n}\neq \mathbf{0}} 
\frac{e^{-|\mathbf{n}| m_{\pi} {\rm L}}}{|\mathbf{n}|} 
\ e^{-i\mathbf{n}\cdot\bm{\phi}}
\ \ \ \ ,
\label{eq:fullsum}
\end{eqnarray}
of which the first few terms are 
\begin{eqnarray}
S({\bm\phi}) & = & 
2\  e^{-m_{\pi} {\rm L}}\ 
\left( \cos\phi_x + \cos\phi_y + \cos\phi_z\right)
\nonumber\\
& + &  
2\sqrt{2}\  e^{-\sqrt{2} m_{\pi} {\rm L}}\ 
\left( \cos\phi_x  \cos\phi_y +  \cos\phi_x  \cos\phi_z +  \cos\phi_y  \cos\phi_z \right)
\nonumber\\
& + &  
{8\over\sqrt{3}}\ e^{-\sqrt{3} m_{\pi} {\rm L}}\ 
\cos\phi_x  \cos\phi_y  \cos\phi_z \ 
\nonumber\\
& + &  
e^{-2 m_{\pi} {\rm L}}\ 
 \left( \cos 2\phi_x + \cos 2\phi_y + \cos 2\phi_z\right)
\ +\ \cdots
\ \ \ \ .
\label{eq:partsum}
\end{eqnarray}
For PBCs, with ${\bm\phi}=(0,0,0)$, the first few terms in the 
sum in Eq.~(\ref{eq:fullsum}) and (\ref{eq:partsum}) are
\begin{eqnarray}
S({\bf 0}) & = & 
6 \ e^{-m_{\pi} {\rm L}}\ 
+
6\sqrt{2} \ e^{-\sqrt{2} m_{\pi} {\rm L}}\ 
+
{8\over\sqrt{3}} \ e^{-\sqrt{3} m_{\pi} {\rm L}}\ 
+
3 \ e^{-2 m_{\pi} {\rm L}}\ 
\ +\ \cdots
\ \ \ \ ,
\label{eq:partsum}
\end{eqnarray}
while for APBs, with ${\bm\phi}=(\pi,\pi,\pi)$, the sum becomes
\begin{eqnarray}
S({\bm \pi}) & = & 
-6 \ e^{-m_{\pi} {\rm L}}\ 
+
6\sqrt{2} \ e^{-\sqrt{2} m_{\pi} {\rm L}}\ 
-
{8\over\sqrt{3}} \ e^{-\sqrt{3} m_{\pi} {\rm L}}\ 
+
3 \ e^{-2 m_{\pi} {\rm L}}\ 
\ -\ \cdots
\ \ \ \ .
\label{eq:partsum}
\end{eqnarray}
It is obvious that the leading terms vanish
in the average, with $(S({\bf 0})+S({\bm \pi}))/2 = 6\sqrt{2}\  e^{-\sqrt{2} m_{\pi} {\rm L}} +...~$.
A particularly interesting twist is ${\bm\phi}=({\pi\over 2},{\pi\over 2},{\pi\over 2})$, 
induced by i-PBCs, 
for which the first three terms in the sum
vanish, leaving 
\begin{eqnarray}
S({{\bm \pi}\over 2})
&  =   & 
-3\  e^{-2 m_{\pi} {\rm L}} 
\ +\ \cdots
\ \ \ \ .
\label{eq:magictwist}
\end{eqnarray}
Finally, twist averaging this function gives
\begin{eqnarray}
\langle S({\bm \phi}) \rangle_{\bm\phi}\ =\ 
\int\ {d^3{\bm\phi}\over (2\pi)^3}\ 
S({\bm \phi})
&  =   & 
0
\ \ \ \ .
\label{eq:twistAvS}
\end{eqnarray}
%

%\end{small}

%\newpage
%\begin{small}
\section{Quantization Conditions \label{app: QC}}
%%%%%%%%%%%%%%%%%%%%%%%%%%%%%%%%%%%%%%
\noindent
The NN FV QCs in the channels
that have an overlap with the $\siii$-$\diii$ coupled channels
are listed in this appendix for a selection of twist angles. 
With the notation of Ref.~\cite{Briceno:2013lba},
the QC for the irrep $\Gamma_i$ can be written as
\begin{eqnarray}
\det\left({\mathbb{M}}^{(\Gamma_i)}
\ +\  i \frac{p^*}{8\pi E^*}-\mathcal{F}^{(\Gamma_i),{\textbf{d},\bm{\phi}_1,\bm{\phi}_2}}\right)=0
\ \ \ ,
\label{QC-simplified}
\end{eqnarray} 
where
\begin{eqnarray}
\mathcal{F}^{(\Gamma_i),{\textbf{d},\bm{\phi}_1,\bm{\phi}_2}}(p^{*2}; {\rm L} )
& = &
\frac{1}{2E^*}\sum_{l,m}\frac{1}{p^{*l}}~{\mathbb{F}
}_{lm}^{(\Gamma_i)}~{c_{lm}^{\textbf{d},\bm{\phi}_1,\bm{\phi}_2}(p^{*2};{\rm L})}
\ \ \ ,
\nonumber\\
{\mathbb{M}}^{(\Gamma_i)}
& = & \left( \mathcal{M}^{-1}\right)_{\Gamma_i}
\ \ \ .
\label{def-F}
\end{eqnarray}
where 
${c_{lm}^{\textbf{d},\bm{\phi}_1,\bm{\phi}_2}(p^{*2};{\rm L})}$ functions are defined in
Eqs.~(\ref{clm}),(\ref{Zlm}) and (\ref{r-TBC}), 
$E^*$ is NN CM energy and $p^*$ is the on-shell momentum of each nucleon in the CM 
frame.~\footnote{
The relativistic normalization of states has been used such that for a single  S-wave channel with phase shift $\delta$, 
the scattering amplitude is
${\cal M} =  {8\pi E^*\over p^*} {\left( e^{2 i \delta}-1\right) \over 2i}$.
} 
In the summation over ``$m$'' in Eq.~(\ref{def-F}), only the 
${\mathbb{F}}_{lm}^{(\Gamma_i)}$ listed below are included as the other contributions have already 
been summed using the symmetries of the systems.
In the following we set $\bm{\phi}_1=-\bm{\phi}_2=\bm{\phi}$.
It is straightforward to decompose $\mathcal{M}^{-1}$ into 
$\left( \mathcal{M}^{-1}\right)_{\Gamma_i}$
using the eigenvectors of the FV functions~\cite{Thomas:2011rh,Dudek:2012gj}.
For notational convenience,
$\mathcal{M}_{J,L}$ denotes the scattering amplitude in the channel with total
angular momentum $J$ 
and orbital angular momentum $L$.
 $\mathcal{M}_{1,SD}$ is the amplitude between $S$ and $D$ partial
waves in the $J=1$ channel, 
and $\text{det} \mathcal{M}_1$ is the determinant of the $J=1$ sector of the scattering-amplitude matrix,
\begin{eqnarray}
\det\mathcal{M}_{1}=\det \left( \begin{array}{cc}
\mathcal{M}_{1,S}&\mathcal{M}_{1,SD}\\
\mathcal{M}_{1,DS}&\mathcal{M}_{1,D}\\
\end{array} \right)
\ \ \ .
\end{eqnarray}
%
%\end{small}

%%%%%%%%%%%%%%%%%%%%%%%%%%%%%%%%%%%%%%
\subsection{${ \phi}=(0,0,0)$}
\begin{footnotesize}
\begin{align}
& \mathbb{T}_1: \hspace{0.4cm}
\mathbb{F}_{00}^{(\mathbb{T}_1)}=\textbf{I}_{3},\hspace{0.15cm}
\mathbb{F}_{40}^{(\mathbb{T}_1)}=
\left(
\begin{array}{ccc}
 0 & 0 & 0 \\
 0 & 0 & \frac{2 \sqrt{6}}{7} \\
 0 & \frac{2 \sqrt{6}}{7} & \frac{2}{7} \\
\end{array}
\right),\hspace{0.15cm}
{\mathbb{M}}^{(\mathbb{T}_1)}=\left(
\begin{array}{ccc}
 \frac{\mathcal{M}_{1,D}}{{\det\mathcal{M}_{1}}} & -\frac{\mathcal{M}_{1,SD}}{\det\mathcal{M}_{1}} & 0 \\
 -\frac{\mathcal{M}_{1,SD}}{\det\mathcal{M}_{1}} & \frac{\mathcal{M}_{1,S}}{\det\mathcal{M}_{1}} & 0 \\
 0 & 0 &\mathcal{M}_{3,D}^{-1} \\
\end{array}
\right)
\ \ \ . 
\label{000T1}
\end{align}
\end{footnotesize}

%%%%%%%%%%%%%%%%%%%%%%%
\subsection{${ \phi}=(\frac{\pi}{2},\frac{\pi}{2},\frac{\pi}{2})$}
\begin{footnotesize}
\begin{align}
& \mathbb{A}_2:\hspace{0.4cm}
\mathbb{F}_{00}^{(\mathbb{A}_2)}=\textbf{I}_{6},\hspace{0.15cm}
\mathbb{F}_{10}^{(\mathbb{A}_2)}= \left(
\begin{array}{cccccc}
 0 & -1 & \sqrt{2} & 0 & 0 & 0 \\
 -1 & 0 & 0 & \sqrt{2} & 0 & 0 \\
 \sqrt{2} & 0 & 0 & -\frac{1}{5} & 0 & 0 \\
 0 & \sqrt{2} & -\frac{1}{5} & 0 & \frac{3 \sqrt{6}}{5} & 0 \\
 0 & 0 & 0 & \frac{3 \sqrt{6}}{5} & 0 & 0 \\
 0 & 0 & 0 & 0 & 0 & 0
\end{array}
\right)
\ ,
\hspace{0.15cm}
\mathbb{F}_{22}^{(\mathbb{A}_2)}
= 
i \times 
\left(
\begin{array}{cccccc}
 0 & 0 & 0 & 2 \sqrt{\frac{3}{5}} & 0 & 0 \\
 0 & 0 & 2 \sqrt{\frac{3}{5}} & 0 & -3  \sqrt{\frac{2}{5}} & 0 \\
 0 & 2 \sqrt{\frac{3}{5}} & - \sqrt{\frac{6}{5}} & 0 & \frac{6 }{7 \sqrt{5}} & 0 \\
 2  \sqrt{\frac{3}{5}} & 0 & 0 & - \sqrt{\frac{6}{5}} & 0 & 0 \\
 0 & -3  \sqrt{\frac{2}{5}} & \frac{6 }{7 \sqrt{5}} & 0 & -\frac{8}{7}  \sqrt{\frac{6}{5}} & 0 \\
 0 & 0 & 0 & 0 & 0 & \frac{2  \sqrt{30}}{7}
\end{array}
\right)
\ ,
\nonumber\\
&\hspace{1.2cm}
\mathbb{F}_{30}^{(\mathbb{A}_2)}=
\left(
\begin{array}{cccccc}
 0 & 0 & 0 & 0 & \frac{2}{\sqrt{7}} & \sqrt{\frac{5}{7}} \\
 0 & 0 & 0 & 0 & 0 & 0 \\
 0 & 0 & 0 & \frac{6 \sqrt{\frac{3}{7}}}{5} & 0 & 0 \\
 0 & 0 & \frac{6 \sqrt{\frac{3}{7}}}{5} & 0 & -\frac{4 \sqrt{\frac{2}{7}}}{5} & \sqrt{\frac{5}{14}} \\
 \frac{2}{\sqrt{7}} & 0 & 0 & -\frac{4 \sqrt{\frac{2}{7}}}{5} & 0 & 0 \\
 \sqrt{\frac{5}{7}} & 0 & 0 & \sqrt{\frac{5}{14}} & 0 & 0
\end{array}
\right)
\ ,
\hspace{0.15cm}
\mathbb{F}_{32}^{(\mathbb{A}_2)}=
i \times 
\left(
\begin{array}{cccccc}
 0 & 0 & 0 & 0 &  \sqrt{\frac{10}{21}} & -2  \sqrt{\frac{2}{21}} \\
 0 & 0 & 0 & 0 & 0 & 0 \\
 0 & 0 & 0 & 3  \sqrt{\frac{2}{35}} & 0 & 0 \\
 0 & 0 & 3 \sqrt{\frac{2}{35}} & 0 & -\frac{4 }{\sqrt{105}} & -\frac{2 }{\sqrt{21}} \\
  \sqrt{\frac{10}{21}} & 0 & 0 & -\frac{4 }{\sqrt{105}} & 0 & 0 \\
 -2  \sqrt{\frac{2}{21}} & 0 & 0 & -\frac{2 }{\sqrt{21}} & 0 & 0
\end{array}
\right)
\ ,
\nonumber
\end{align}
\begin{align}
&\hspace{-0.4cm}
\mathbb{F}_{40}^{(\mathbb{A}_2)}=
\left(
\begin{array}{cccccc}
 0 & 0 & 0 & 0 & 0 & 0 \\
 0 & 0 & 0 & 0 & 0 & 0 \\
 0 & 0 & 0 & 0 & \frac{4 \sqrt{\frac{2}{3}}}{7} & \frac{2 \sqrt{\frac{10}{3}}}{7} \\
 0 & 0 & 0 & 0 & 0 & 0 \\
 0 & 0 & \frac{4 \sqrt{\frac{2}{3}}}{7} & 0 & -\frac{4}{21} & \frac{4 \sqrt{5}}{21} \\
 0 & 0 & \frac{2 \sqrt{\frac{10}{3}}}{7} & 0 & \frac{4 \sqrt{5}}{21} & -\frac{2}{21}
\end{array}
\right)\ ,
\hspace{0.4cm}
\mathbb{F}_{42}^{(\mathbb{A}_2)}=
i \times 
\left(
\begin{array}{cccccc}
 0 & 0 & 0 & 0 & 0 & 0 \\
 0 & 0 & 0 & 0 & 0 & 0 \\
 0 & 0 & 0 & 0 & \frac{8}{7}  \sqrt{\frac{5}{3}} & -\frac{1}{\sqrt{3}} \\
 0 & 0 & 0 & 0 & 0 & 0 \\
 0 & 0 & \frac{8}{7}  \sqrt{\frac{5}{3}} & 0 & -\frac{4  \sqrt{10}}{21} & -\frac{ \sqrt{2}}{3} \\
 0 & 0 & -\frac{1}{\sqrt{3}} & 0 & -\frac{ \sqrt{2}}{3} & -\frac{2  \sqrt{10}}{21}
\end{array}
\right)
\ ,
\nonumber\\
&\hspace{-0.4cm}{\mathbb{M}}^{(\mathbb{A}_2)}=
%%%%%%%%%%%%%%%%%%
\left(
\begin{array}{cccccc}
 \mathcal{M}_{0,P}^{-1} & 0 & 0 & 0 & 0 & 0 \\
 0 & \frac{\mathcal{M}_{1,D}}{{\det\mathcal{M}_{1}}} & -\frac{\mathcal{M}_{1,SD}}{{\det\mathcal{M}_{1}}} & 0 & 0 & 0 \\
 0 & -\frac{\mathcal{M}_{1,SD}}{{\det\mathcal{M}_{1}}} & \frac{\mathcal{M}_{1,S}}{{\det\mathcal{M}_{1}}} & 0 & 0 & 0 \\
 0 & 0 & 0 & \mathcal{M}_{2,P}^{-1} & 0 & 0 \\
 0 & 0 & 0 & 0 & \mathcal{M}_{3,D}^{-1} & 0 \\
 0 & 0 & 0 & 0 & 0 & \mathcal{M}_{3,D}^{-1}
\end{array}
\right)
\ .
\label{pi2pi2pi2A2}
\end{align}
\begin{align}
& \mathbb{E}:\hspace{0.4cm}\mathbb{F}_{00}^{(\mathbb{E})}=\textbf{I}_{9} ,\hspace{0.15cm}
\mathbb{F}_{10}^{(\mathbb{E})}=
\left(
\begin{array}{ccccccccc}
 0 & \sqrt{\frac{3}{2}} & 0 & 0 & \sqrt{\frac{3}{2}} & 0 & 0 & 0 & 0 \\
 \sqrt{\frac{3}{2}} & 0 & \frac{\sqrt{3}}{2} & 0 & 0 & 0 & \frac{3 \sqrt{\frac{3}{5}}}{2} & 0 & 0 \\
 0 & \frac{\sqrt{3}}{2} & 0 & 0 & -\frac{\sqrt{3}}{10} & 0 & 0 & 0 & 0 \\
 0 & 0 & 0 & 0 & 0 & \sqrt{\frac{3}{5}} & 0 & \sqrt{\frac{6}{5}} & 0 \\
 \sqrt{\frac{3}{2}} & 0 & -\frac{\sqrt{3}}{10} & 0 & 0 & 0 & \frac{\sqrt{\frac{3}{5}}}{2} & 0 & \frac{4 \sqrt{3}}{5} \\
 0 & 0 & 0 & \sqrt{\frac{3}{5}} & 0 & 0 & 0 & 0 & 0 \\
 0 & \frac{3 \sqrt{\frac{3}{5}}}{2} & 0 & 0 & \frac{\sqrt{\frac{3}{5}}}{2} & 0 & 0 & 0 & 0 \\
 0 & 0 & 0 & \sqrt{\frac{6}{5}} & 0 & 0 & 0 & 0 & 0 \\
 0 & 0 & 0 & 0 & \frac{4 \sqrt{3}}{5} & 0 & 0 & 0 & 0
\end{array}
\right)
\ ,
\nonumber\\
& \hspace{1.2cm}\mathbb{F}_{22}^{(\mathbb{E})}=
i \times 
\left(
\begin{array}{ccccccccc}
 0 & 0 & - \sqrt{\frac{3}{5}} & 0 & 0 & 0 & -\sqrt{3} & 0 & -2 \sqrt{\frac{3}{5}} \\
 0 & - \sqrt{\frac{3}{10}} & 0 & 0 & -3  \sqrt{\frac{3}{10}} & 0 & 0 & 0 & 0 \\
 - \sqrt{\frac{3}{5}} & 0 &  \sqrt{\frac{3}{10}} & 0 & 0 & 0 & - \sqrt{\frac{3}{2}} & 0 & \frac{2}{7}  \sqrt{\frac{6}{5}} \\
 0 & 0 & 0 &  \sqrt{\frac{6}{5}} & 0 & 0 & 0 & 0 & 0 \\
 0 & -3  \sqrt{\frac{3}{10}} & 0 & 0 & - \sqrt{\frac{3}{10}} & 0 & 0 & 0 & 0 \\
 0 & 0 & 0 & 0 & 0 & \frac{ \sqrt{30}}{7} & 0 & -\frac{2 \sqrt{15}}{7} & 0 \\
 -\sqrt{3} & 0 & - \sqrt{\frac{3}{2}} & 0 & 0 & 0 & -\frac{1}{7}  \sqrt{\frac{15}{2}} & 0 & -\frac{2  \sqrt{6}}{7} \\
 0 & 0 & 0 & 0 & 0 & -\frac{2 \sqrt{15}}{7} & 0 & 0 & 0 \\
 -2  \sqrt{\frac{3}{5}} & 0 & \frac{2}{7}  \sqrt{\frac{6}{5}} & 0 & 0 & 0 & -\frac{2  \sqrt{6}}{7} & 0 & -\frac{6}{7}  \sqrt{\frac{6}{5}}
\end{array}
\right)
\ ,
\nonumber\\
& \hspace{1.2cm}\mathbb{F}_{30}^{(\mathbb{E})}=
\left(
\begin{array}{ccccccccc}
 0 & 0 & 0 & 0 & 0 & 0 & 0 & 0 & 0 \\
 0 & 0 & 0 & 0 & 0 & -\sqrt{\frac{5}{14}} & -\frac{2}{\sqrt{35}} & \frac{\sqrt{\frac{5}{7}}}{2} & -\frac{2}{\sqrt{7}} \\
 0 & 0 & 0 & \frac{3}{\sqrt{14}} & -\frac{6}{5 \sqrt{7}} & 0 & 0 & 0 & 0 \\
 0 & 0 & \frac{3}{\sqrt{14}} & 0 & 0 & \frac{2}{\sqrt{35}} & -\sqrt{\frac{5}{14}} & 2 \sqrt{\frac{2}{35}} & \frac{1}{\sqrt{14}} \\
 0 & 0 & -\frac{6}{5 \sqrt{7}} & 0 & 0 & \sqrt{\frac{5}{14}} & -\frac{4}{\sqrt{35}} & -\frac{\sqrt{\frac{5}{7}}}{2} & -\frac{2}{5 \sqrt{7}} \\
 0 & -\sqrt{\frac{5}{14}} & 0 & \frac{2}{\sqrt{35}} & \sqrt{\frac{5}{14}} & 0 & 0 & 0 & 0 \\
 0 & -\frac{2}{\sqrt{35}} & 0 & -\sqrt{\frac{5}{14}} & -\frac{4}{\sqrt{35}} & 0 & 0 & 0 & 0 \\
 0 & \frac{\sqrt{\frac{5}{7}}}{2} & 0 & 2 \sqrt{\frac{2}{35}} & -\frac{\sqrt{\frac{5}{7}}}{2} & 0 & 0 & 0 & 0 \\
 0 & -\frac{2}{\sqrt{7}} & 0 & \frac{1}{\sqrt{14}} & -\frac{2}{5 \sqrt{7}} & 0 & 0 & 0 & 0
\end{array}
\right)
\ ,
\nonumber
\end{align}
\begin{align}
& \hspace{0.2cm}\mathbb{F}_{32}^{(\mathbb{E})}=
i \times 
\left(
\begin{array}{ccccccccc}
 0 & 0 & 0 & 0 & 0 & 0 & 0 & 0 & 0 \\
 0 & 0 & 0 & 0 & 0 & \frac{2 }{\sqrt{21}} & -\sqrt{\frac{2}{21}} & -\sqrt{\frac{2}{21}} & -\sqrt{\frac{10}{21}} \\
 0 & 0 & 0 & -2 \sqrt{\frac{3}{35}} & -\sqrt{\frac{6}{35}} & 0 & 0 & 0 & 0 \\
 0 & 0 & -2  \sqrt{\frac{3}{35}} & 0 & 0 &  \sqrt{\frac{2}{21}} & \frac{2 }{\sqrt{21}} & \frac{2 }{\sqrt{21}} & -\frac{2 }{\sqrt{105}} \\
 0 & 0 & -\sqrt{\frac{6}{35}} & 0 & 0 & -\frac{2 }{\sqrt{21}} & -2 \sqrt{\frac{2}{21}} &  \sqrt{\frac{2}{21}} & - \sqrt{\frac{2}{105}} \\
 0 & \frac{2 }{\sqrt{21}} & 0 &  \sqrt{\frac{2}{21}} & -\frac{2 }{\sqrt{21}} & 0 & 0 & 0 & 0 \\
 0 & - \sqrt{\frac{2}{21}} & 0 & \frac{2 }{\sqrt{21}} & -2  \sqrt{\frac{2}{21}} & 0 & 0 & 0 & 0 \\
 0 & - \sqrt{\frac{2}{21}} & 0 & \frac{2 }{\sqrt{21}} &  \sqrt{\frac{2}{21}} & 0 & 0 & 0 & 0 \\
 0 & - \sqrt{\frac{10}{21}} & 0 & -\frac{2 }{\sqrt{105}} & - \sqrt{\frac{2}{105}} & 0 & 0 & 0 & 0
\end{array}
\right)
\ ,
\nonumber\\
& \hspace{0.2cm}\mathbb{F}_{40}^{(\mathbb{E})}=
\left(
\begin{array}{ccccccccc}
 0 & 0 & 0 & 0 & 0 & 0 & 0 & 0 & 0 \\
 0 & 0 & 0 & 0 & 0 & 0 & 0 & 0 & 0 \\
 0 & 0 & 0 & 0 & 0 & 0 & 0 & \frac{2 \sqrt{5}}{7} & -\frac{2}{7} \\
 0 & 0 & 0 & 0 & 0 & 0 & 0 & 0 & 0 \\
 0 & 0 & 0 & 0 & 0 & 0 & 0 & 0 & 0 \\
 0 & 0 & 0 & 0 & 0 & \frac{4}{63} & -\frac{20 \sqrt{2}}{63} & \frac{10 \sqrt{2}}{63} & \frac{10 \sqrt{10}}{63} \\
 0 & 0 & 0 & 0 & 0 & -\frac{20 \sqrt{2}}{63} & -\frac{16}{63} & -\frac{10}{63} & -\frac{10 \sqrt{5}}{63} \\
 0 & 0 & \frac{2 \sqrt{5}}{7} & 0 & 0 & \frac{10 \sqrt{2}}{63} & -\frac{10}{63} & \frac{2}{9} & -\frac{4 \sqrt{5}}{63} \\
 0 & 0 & -\frac{2}{7} & 0 & 0 & \frac{10 \sqrt{10}}{63} & -\frac{10 \sqrt{5}}{63} & -\frac{4 \sqrt{5}}{63} & -\frac{2}{63}
\end{array}
\right)
\ ,
\hspace{0.15cm}\mathbb{F}_{42}^{(\mathbb{E})}=
i \times 
\left(
\begin{array}{ccccccccc}
 0 & 0 & 0 & 0 & 0 & 0 & 0 & 0 & 0 \\
 0 & 0 & 0 & 0 & 0 & 0 & 0 & 0 & 0 \\
 0 & 0 & 0 & 0 & 0 & 0 & 0 & -\frac{1}{\sqrt{2}} & -\frac{2  \sqrt{10}}{7} \\
 0 & 0 & 0 & 0 & 0 & 0 & 0 & 0 & 0 \\
 0 & 0 & 0 & 0 & 0 & 0 & 0 & 0 & 0 \\
 0 & 0 & 0 & 0 & 0 & \frac{4  \sqrt{10}}{63} & \frac{2  \sqrt{5}}{9} & \frac{20  \sqrt{5}}{63} & -\frac{5 }{9} \\
 0 & 0 & 0 & 0 & 0 & \frac{2 \sqrt{5}}{9} & -\frac{16 \sqrt{10}}{63} & \frac{1}{9}  \sqrt{\frac{5}{2}} & -\frac{50 \sqrt{2}}{63} \\
 0 & 0 & -\frac{1}{\sqrt{2}} & 0 & 0 & \frac{20  \sqrt{5}}{63} & \frac{1}{9} \sqrt{\frac{5}{2}} & \frac{2  \sqrt{10}}{9} & \frac{\sqrt{2}}{9} \\
 0 & 0 & -\frac{2  \sqrt{10}}{7} & 0 & 0 & -\frac{5 }{9} & -\frac{50  \sqrt{2}}{63} & \frac{ \sqrt{2}}{9} & -\frac{2 \sqrt{10}}{63}
\end{array}
\right)
\ ,
\nonumber\\
&\hspace{0.2cm}{\mathbb{M}}^{(\mathbb{E})}=\left(
\begin{array}{ccccccccc}
 \frac{\mathcal{M}_{1,D}}{{\det\mathcal{M}_{1}}} & 0 & - \frac{\mathcal{M}_{1,SD}}{{\det\mathcal{M}_{1}}} & 0 & 0 & 0 & 0 & 0 & 0 \\
 0 & \mathcal{M}_{1,P}^{-1} & 0 & 0 & 0 & 0 & 0 & 0 & 0 \\
 - \frac{\mathcal{M}_{1,SD}}{{\det\mathcal{M}_{1}}} & 0 &  \frac{\mathcal{M}_{1,S}}{{\det\mathcal{M}_{1}}} & 0 & 0 & 0 & 0 & 0 & 0 \\
 0 & 0 & 0 & \mathcal{M}_{2,P}^{-1} & 0 & 0 & 0 & 0 & 0 \\
 0 & 0 & 0 & 0 & \mathcal{M}_{2,P}^{-1} & 0 & 0 & 0 & 0 \\
 0 & 0 & 0 & 0 & 0 & \mathcal{M}_{2,D}^{-1} & 0 & 0 & 0 \\
 0 & 0 & 0 & 0 & 0 & 0 & \mathcal{M}_{2,D}^{-1} & 0 & 0 \\
 0 & 0 & 0 & 0 & 0 & 0 & 0 & \mathcal{M}_{3,D}^{-1} & 0 \\
 0 & 0 & 0 & 0 & 0 & 0 & 0 & 0 & \mathcal{M}_{3,D}^{-1}
\end{array}
\right)
\ .
\label{pi2pi2pi2E}
\end{align}
\end{footnotesize}

%%%%%%%%%%%%%%%%%%%%%%%
\subsection{${ \phi}=(\pi,\pi,\pi)$}
\begin{footnotesize}
\begin{align}
& \mathbb{A}_2:\hspace{0.4cm}
\mathbb{F}_{00}^{(\mathbb{A}_2)}=\textbf{I}_{4} ,\hspace{0.15cm}
\mathbb{F}_{40}^{(\mathbb{A}_2)}=
\left(
\begin{array}{cccc}
 0 & 0 & 0 & 0 \\
 0 & 0 & \frac{2 \sqrt{6}}{7} & 0 \\
 0 & \frac{2 \sqrt{6}}{7} & \frac{2}{7} & 0 \\
 0 & 0 & 0 & -\frac{4}{7}
\end{array}
\right), \hspace{0.15cm}
{\mathbb{M}}^{(\mathbb{A}_2)}=\left(
\begin{array}{cccc}
 \frac{\mathcal{M}_{1,D}}{{\det\mathcal{M}_{1}}} & -\frac{\mathcal{M}_{1,SD}}{\det\mathcal{M}_{1}} & 0 & 0 \\
 -\frac{\mathcal{M}_{1,SD}}{\det\mathcal{M}_{1}} & \frac{\mathcal{M}_{1,S}}{\det\mathcal{M}_{1}} & 0 & 0 \\
 0 & 0 &\mathcal{M}_{3,D}^{-1} & 0 \\
 0 & 0 & 0 & \mathcal{M}_{3,D}^{-1} \\
\end{array}
\right)
\ .
\label{pipipiA2}
\\
& \mathbb{E}:\hspace{0.4cm}\mathbb{F}_{00}^{(\mathbb{E})}=\textbf{I}_{6},\hspace{0.15cm}
\mathbb{F}_{40}^{(\mathbb{E})}=
\left(
\begin{array}{cccccc}
 0 & 0 & 0 & 0 & 0 & 0 \\
 0 & 0 & \frac{2 \sqrt{6}}{7} & 0 & 0 & 0 \\
 0 & \frac{2 \sqrt{6}}{7} & \frac{2}{7} & 0 & 0 & 0 \\
 0 & 0 & 0 & \frac{8}{21} & -\frac{10 \sqrt{2}}{21} & 0 \\
 0 & 0 & 0 & -\frac{10 \sqrt{2}}{21} & -\frac{2}{21} & 0 \\
 0 & 0 & 0 & 0 & 0 & -\frac{4}{7}
\end{array}
\right)
\ ,
\hspace{0.15cm}{\mathbb{M}}^{(\mathbb{E})}=
 \left(
\begin{array}{cccccc}
  \frac{\mathcal{M}_{1,D}}{{\det\mathcal{M}_{1}}} & -\frac{\mathcal{M}_{1,SD}}{\det\mathcal{M}_{1}} & 0 & 0 & 0 & 0 \\
-\frac{\mathcal{M}_{1,SD}}{\det\mathcal{M}_{1}} & \frac{\mathcal{M}_{1,S}}{\det\mathcal{M}_{1}} & 0 & 0 & 0 & 0 \\
 0 & 0 & \mathcal{M}_{3,D}^{-1} & 0 & 0 & 0 \\
 0 & 0 & 0 & \mathcal{M}_{2,D}^{-1} & 0 & 0 \\
 0 & 0 & 0 & 0 & \mathcal{M}_{3,D}^{-1} & 0 \\
 0 & 0 & 0 & 0 & 0 & \mathcal{M}_{2,D}^{-1} \\
\end{array}
\right)
\ .
\label{pipipiE}
\end{align}
\end{footnotesize}

%%%%%%%%%%%%%%%%%%%%%%%%%%%%%%%%%%%%%%

%\newpage
%\begin{small}
\section{Twisted $c_{lm}^{ {\bf d},{\bf\phi}_1,{\bf\phi}_2}$  Functions for Systems at Rest
\label{app:TwistC}}
%%%%%%%%%%%%%%%%%%%%%%%%%%%%%%%%%%%%%%
\noindent
To understand the relative contributions of phase shifts beyond the $\alpha$ wave to the 
deuteron binding energy, it is helpful to 
consider the expansions of the $c_{lm}^{ {\bf d},{\bf\phi}_1,{\bf\phi}_2}$  functions.
As i-PBCs, with the twist angles $\bm{\phi}=(\frac{\pi}{2},\frac{\pi}{2},\frac{\pi}{2})$, lead to the most significant reduction 
in the FV corrections, we focus on these angles in the expansions, restricting ourselves to systems at rest.
The general form of the $c_{lm}^{ {\bf d},{\bm\phi}_1,{\bm\phi}_2}$  functions for $\mathbf{d}=\mathbf{0}$ and ${\bm\phi}_1=-{\bm\phi}_2=\bm\phi$ is
\begin{eqnarray}
{c_{lm}^{\textbf{0},\bm{\phi},-\bm{\phi}}(-\kappa^2;{\rm L})}
& = & 
\frac{i^l}{\pi^{3/2}}\sum_{\mathbf{n} \neq \mathbf{0}}
\ e^{-i \mathbf{n} \cdot \bm{\phi}}\ 
Y_{lm}(\hat{\mathbf{n}})
\int_{0}^{\infty}dk~\frac{k^{l+2}}{k^2+\kappa^2}~j_l(n k {\rm L})
\ \ \ ,
\end{eqnarray}
where $n=|{\bf n}|$.
By direct evaluation of the integral,
it is straightforward to show that
\begin{eqnarray}
&&
{c_{00}^{\textbf{0},\bm{\phi},-\bm{\phi}}(-\kappa^2;{\rm L})}
\ =\ 
-{\kappa\over 4\pi}\ +\ 
 \sqrt{4\pi}~
\sum_{\mathbf{n} \neq \mathbf{0}}
\ e^{-i \mathbf{n} \cdot \bm{\phi}}\ 
Y_{00}(\hat{\mathbf{n}})~
\frac{e^{-n \kappa {\rm L}}}{4\pi n {\rm L}}
\ \ ,
\\
&&
{c_{1m}^{\textbf{0},\bm{\phi},-\bm{\phi}}(-\kappa^2;{\rm L})}=
(i\kappa)\sqrt{4\pi}~
\sum_{\mathbf{n} \neq \mathbf{0}}
\ e^{-i \mathbf{n} \cdot \bm{\phi}}\ 
Y_{1m}(\hat{\mathbf{n}})~\left(1+\frac{1}{n \kappa {\rm L}}\right)~
\frac{e^{-n \kappa {\rm L}}}{4\pi n {\rm L}}
\ \ ,
\\
&&{c_{2m}^{\textbf{0},\bm{\phi},-\bm{\phi}}(-\kappa^2;{\rm L})}=
(i\kappa)^2\sqrt{4\pi}~
\sum_{\mathbf{n} \neq \mathbf{0}}
\ e^{-i \mathbf{n} \cdot \bm{\phi}}\ 
Y_{2m}(\hat{\mathbf{n}})~
\left(1+\frac{3}{n \kappa {\rm L}}+\frac{3}{n^2 \kappa^2 {\rm L}^2}\right)~
\frac{e^{-n \kappa {\rm L}}}{4\pi n {\rm L}}
\ \ ,
\\
&&{c_{3m}^{\textbf{0},\bm{\phi},-\bm{\phi}}(-\kappa^2;{\rm L})}=
(i\kappa)^3\sqrt{4\pi}~
\sum_{\mathbf{n} \neq \mathbf{0}}
\ e^{-i \mathbf{n} \cdot \bm{\phi}}\ 
Y_{3m}(\hat{\mathbf{n}})~
\left(1+\frac{6}{n \kappa {\rm L}}+\frac{15}{n^2 \kappa^2 {\rm L}^2}+\frac{15}{n^3 \kappa^3 {\rm L}^3}\right)~
\frac{e^{-n \kappa {\rm L}}}{4\pi n {\rm L}}
\ \ ,
\\
&&{c_{4m}^{\textbf{0},\bm{\phi},-\bm{\phi}}(-\kappa^2;{\rm L})}=
(i\kappa)^4\sqrt{4\pi}~
\sum_{\mathbf{n} \neq \mathbf{0}}
\ e^{-i \mathbf{n} \cdot \bm{\phi}}\ 
Y_{4m}(\hat{\mathbf{n}})~
\left(1+\frac{10}{n\kappa {\rm L}} 
+\frac{45}{n^2\kappa^{2}{\rm L}^2}+\frac{105}{n^3\kappa^{3}{\rm L}^3}+\frac{105}{n^4\kappa^{4}{\rm L}^4}\right)~
\frac{e^{-n\kappa {\rm L}}}{4\pi n{\rm L}}
\ \ .
\nonumber\\
\end{eqnarray}
These functions are of the form
\begin{eqnarray}
F_{lm} & = & 
\sum_{\mathbf{n} \neq \mathbf{0}}\ e^{-i \mathbf{n} \cdot \bm{\phi}}\ 
Y_{1m}(\hat{\mathbf{n}})~f(n)
\nonumber\\
& = & \alpha^{(1)}_{lm} \ f(1)\ +\ \alpha^{(\sqrt{2})}_{lm}\  f(\sqrt{2})\ +\ \alpha^{(\sqrt{3})}_{lm}\  f(\sqrt{3})\ +\ \alpha^{(2)}_{lm} f(2)
\ +\ \cdots
\ \ \ .
\label{eq:Flm}
\end{eqnarray}
The independent and non-vanishing coefficients $\alpha^{(n)}$ are presented in Table~\ref{tab:alphas} for the twist angles 
$\bm{\phi}=(\frac{\pi}{2},\frac{\pi}{2},\frac{\pi}{2})$.
\begin{center}
\begin{table}[ht!]
\renewcommand{\arraystretch}{1.6}
\scalebox{1.1}{
\begin{tabular}{|c||c|c|c|c|}
\hline 
$(l,m)$ & $\alpha^{(1)}_{lm}$ & $\alpha^{(\sqrt{2})}_{lm}$ & $\alpha^{(\sqrt{3})}_{lm}$ & $\alpha^{(2)}_{lm}$\tabularnewline
\hline 
\hline 
$(0,0)$ & $0$ & $0$ & $0$ & $-\frac{3}{\sqrt{\pi}}$\tabularnewline
\hline
$(1,0)$ & $-i\sqrt{\frac{3}{\pi}}$ & $0$ & $0$ & $0$\tabularnewline
\hline 
$(2,2)$ & $0$ & $-i\sqrt{\frac{15}{2\pi}}$ & $0$ & $0$\tabularnewline
\hline 
$(3,0)$ & $-i\sqrt{\frac{7}{\pi}}$ & $0$ & $0$ & $0$\tabularnewline
\hline 
$(3,2)$ & $0$ & $0$ & $-\frac{2}{3}\sqrt{\frac{70}{\pi}}$ & $0$\tabularnewline
\hline 
$(4,0)$ & $0$ & $0$ & $0$ & $-\frac{21}{4\sqrt{\pi}}$\tabularnewline
\hline 
$(4,2)$ & $0$ & $i\frac{3}{2}\sqrt{\frac{5}{2\pi}}$ & $0$ & $0$\tabularnewline
\hline 
\end{tabular}
}
\caption{Coefficients of independent, non-vanishing terms in the expansion of $F_{lm}$ given in Eq.~(\protect\ref{eq:Flm})}
\label{tab:alphas}
\end{table}
\end{center}
The remaining coefficients are dictated by  symmetry,
\begin{eqnarray}
&&F_{1\pm 1}\ = \ \mp e^{\pm i \pi/4}\ F_{10}
\ \ \ ,
\nonumber\\
&& F_{2+2} \ = \  -F_{2-2} \ =\ -{1\over\sqrt{2}}\ e^{\pm i \pi/4}\ F_{2\pm 1}
\ \ \ ,
\nonumber\\
&& F_{30} \ = \ \mp {4\over\sqrt{10}} e^{\pm i \pi/4} \ F_{3\pm 3}
\ =\  
\pm {4\over\sqrt{6}} e^{\mp i \pi/4} \ F_{3\pm 1}
\ \ ,\ \ 
F_{3-2}\ =\ -F_{3+2}
\ \ \ ,
\nonumber\\
&& F_{4 +2} \ = \ -F_{4 -2} \ =\ 
-{2\over\sqrt{7}}  e^{\mp i \pi/4} \ F_{4\pm 3} 
\ =\  2  e^{\pm i \pi/4} \ F_{4\pm 1} 
\ \ ,\ \ 
F_{40}\ =\ {\sqrt{14\over 5}}\ F_{4\pm4}
\ \ \ .
\label{eq:alpharels}
\end{eqnarray}
The coefficients presented in Table~\ref{tab:alphas} and Eq.~(\ref{eq:alpharels}) show that the leading volume dependences of the 
$c_{lm}^{ {\bf 0},{\bm\phi},-{\bm\phi}}$  functions for i-PBCs are
$c_{00} =-\frac{\kappa}{4\pi}+\mathcal{O}({e^{-2\kappa {\rm L}}}/{{\rm L}})$,
$c_{10} =\mathcal{O}({e^{-\kappa {\rm L}}}/{{\rm L}})$,
$c_{22} =\mathcal{O}({e^{-\sqrt{2}\kappa {\rm L}}}/{{\rm L}})$,
$c_{30} =\mathcal{O}({e^{-\kappa {\rm L}}}/{{\rm L}})$,
$c_{32} =\mathcal{O}({e^{-\sqrt{3}\kappa {\rm L}}}/{{\rm L}})$,
$c_{40} =\mathcal{O}({e^{-2\kappa {\rm L}}}/{{\rm L}})$
and
$c_{42} =\mathcal{O}({e^{-\sqrt{2}\kappa {\rm L}}}/{{\rm L}})$.
As the P-wave contribution to the FV spectra is due to non-zero $c_{1m}$ and $c_{3m}$ functions,
they provide the dominant corrections to the approximate QC in Eq.~(\ref{eq:QC3by3}).

A numerical comparison between these expansions and an exact evaluation of the $c_{lm}^{ {\bf d},{\bm\phi}_1,{\bm\phi}_2}$ functions
reveals that the expansions are only slowly convergent~\cite{Briceno:2013bda}.
Precisions extractions of the energy eigenvalues require the use of the exact evaluations, even in modest volumes.

%\end{small}

\end{document}